\documentclass[%
 reprint, 
 onecolumn,
 amsmath,amssymb,
 aps, physrev,
]{revtex4-2}

\usepackage{graphicx}
\usepackage{dcolumn}
\usepackage{bm}
\usepackage{xcolor}
\renewcommand{\vec}[1]{\boldsymbol{#1}}

\begin{document}

\preprint{APS/123-QED}

\title{\textbf{Laminarising turbulent pipe flow by linear and nonlinear optimisation} 
}%

\author{Shijun Chu}
 \email{schu3@sheffield.ac.uk}
\affiliation{%
 Applied Mathematics, School of Mathematical and Physical Sciences, University of Sheffield, Sheffield S3 7RH, UK.
}%
\author{Ashley P. Willis}
\affiliation{
 Applied Mathematics, School of Mathematical and Physical Sciences, University of Sheffield, Sheffield S3 7RH, UK.
}%
\author{Elena Marensi}
\affiliation{%
School of Mechanical, Aerospace and Civil Engineering, University of Sheffield, Sheffield S1 3JD, UK.
}%


\date{\today}

\begin{abstract}
It has been observed that flattening the mean velocity profile of pipe flow can laminarise turbulence, a promising means to reduce frictional drag substantially.
In the experiments of K\"uhnen $\textit{et al.}$ \cite{kuhnen2018destabilizing},
the flattened profile was achieved by counteracting the flow in the centre and/or 
accelerating the flow near the wall.
In numerical experiments seeking laminarisation, Marensi $\textit{et al.,}$ \cite{marensi2020designing} used a Lagrangian technique to optimise a passive (purely damping) force, where, in contrast to K\"uhnen $\textit{et al.,}$ \citep{kuhnen2018destabilizing} the damping force was found to target turbulent production, mostly near the wall.
To explore whether there is a more efficient body force to eliminate turbulence, we
consider time-independent
active body forces (which may accelerate the flow in a subdomain) with varying spatial dependencies. 
Results confirm that when using an
active force, a
flattened forced laminar profile is needed to eliminate turbulence,
and that
a purely streamwise body force is best for laminarisation. While these results required an expensive nonlinear optimisation, it was also observed that
the optimal forced profile exhibits reduced linear transient growth (TG). To determine whether the reduction of linear TG alone is a sufficient target for the laminarisation of turbulence, a linear Lagrange Multiplier technique is used to minimise TG of perturbations to the forced laminar profile. The optimal velocity profiles reveal that TG for each azimuthal wavenumber strongly depends on the radial velocity gradient at some specific radial interval. The optimal velocity profile obtained by minimising TG of perturbations of azimuthal wavenumber $m=1$ is shown to be able to eliminate turbulence at $Re=2400$, but a more effective reduction of TG, including perturbations of higher $m$, is needed for laminarisation at higher Reynolds number $Re=3000$. The streaks formed with the flattened forced laminar profile reveal the mechanism of laminarisation: it is found that the lift-up mechanism in the more flattened forced laminar profile creates a lower streak (i.e.\ closer to the wall) that is more stable \citep{schoppa2002coherent}, 
so that turbulence is harder to maintain through streak instability. 
Further numerical experiments by disturbing a periodic orbit validated that the breakdown of  turbulence self-sustaining mechanisms is mainly caused by suppression of the formation of streaks.

\end{abstract}

\maketitle



\section{Introduction}
\label{sec:headings}

 In pipe flow, the presence of turbulence causes much higher frictional drag, compared to laminar flow, therefore strategies to reduce or even eliminate shear turbulence are of great interest for many engineering applications.  The development of turbulence control requires an understanding of the dynamics of turbulence, about which research  has been ongoing for over a century since Reynolds $\textit{et al.}$ \cite{reynolds1883xxix}. Some remarkable progress has been made in this field, such as linear stability and transient growth (TG) \citep{schmid2002stability}, analysis of coherent structures \citep{robinson1991coherent}, the self-sustaining process (SSP) \citep{hamilton1995regeneration,jimenez1999autonomous,waleffe1997self}, travelling waves solutions \citep{nagata1990three,faisst2003traveling}, etc. These findings have greatly enhanced our understanding of transitional turbulence, providing avenues for the exploration of turbulence control. 
     
An important step came in the identification of 
the nonlinear vortex-wave interaction \citep{hall1991strongly,hall2018vortex} and the SSP
in shear turbulence \citep{hamilton1995regeneration,waleffe1997self}, which includes the formation of streaks by streamwise vortices, breakdown of the streaks, and regeneration of the streamwise vortices. Subsequently, it was found that an SSP exists which is local to the near-wall region and does not depend on the outer flow \citep{jimenez1999autonomous}. In terms of the mechanism of the SSP, it is widely accepted that the streaks are created in a linear process, which involves the redistribution of the fluid near the wall by streamwise vortices leading to the formation of streaks through the lift-up mechanism \citep{lee1990structure,jimenez2012cascades}. However, there remains some debate about the regeneration of streamwise vortices.  One possibility is that the streamwise vortices are caused by the inflectional instability of streaks \citep{hamilton1995regeneration,jimenez1999autonomous}, while the parent-offspring scenario, whereby an opposite-signed offspring vortex forms immediately
underneath a parent vortex, also gained some support \citep{brooke1993origin,bernard1993vortex}.
Schoppa $\textit{et al.,} $ \cite{schoppa2002coherent} suggested that streamwise vortices are generated from the more numerous normal-mode-stable streaks, via a new streak TG (secondary TG).  However, the streaks break down rapidly and are hard to observe, and it is difficult to distinguish between streak TG and modal streak breakdown as they almost emerge simultaneously during the streak breakdown \citep{cassinelli2017streak}. Recently, Lozano $\textit{et al.,}$ \cite{lozano2021cause} found that TG alone is capable of maintaining wall turbulence with realistic mean velocity and turbulence intensities in the absence of exponential instability, neutral modes and parametric instabilities.
 
Understanding of the SSP has enabled  the development of many strategies to control turbulence for drag reduction. As bursting caused by streamwise vortices is thought to play a dominating role in producing frictional drag \citep{wallace1972wall,lu1973measurements,jimenez2018coherent}, a series of studies attempted to reduce drag  by weakening streamwise vortices. Choi $\textit{et al.,}$ \cite{choi1994active} first achieved this by modelling blowing and suction at the wall, to counteract the streamwise vortices. However, since the vortices are relatively short in the streamwise direction, detection and control would have to be fast and local, which makes this approach impractical. An alternative strategy is to weaken or decorrelate the streaks, which are longer and might be easier to detect and act upon \citep{jimenez1999autonomous}. Schoppa $\textit{et al.,}$ \cite{schoppa1998large} adopted injection from the wall in the experiment and forcing in simulation to stabilize low-speed streaks, achieving drag reduction of up to 30\% and 20\% respectively. Du $\textit{et al.,}$ \cite{du2000suppressing} found that a transverse travelling wave, which is induced by a spanwise force, can eliminate streaks. They used an array of electromagnetic tiles distributed on the wall to produce the spanwise force and obtained net energy savings. It is also possible to reduce drag by forcing the formation of large-scale streaks in pipe flow, which can lead to up to 11\% net power saving \citep{willis2010optimally}. 
Marusic $\textit{et al.,}$ \cite{marusic2021energy} also found that 
oscillations of the wall at frequencies that
target the large-scales can produce drag reduction
at large flow rates.
  Recent numerical and experimental studies suggest that drag reduction can be  realised by applying uniform blowing on the pressure side of an airfoil \citep{atzori2021uniform,miura2022drag}. 
  These methods can achieve significant drag reduction, but the energy saving is more modest when considering the energy input for flow control.
  Other interesting drag reduction methods include the use of superhydrophobic surfaces, whose  effect can be characterised by a slip length in the tangential boundary conditions.  A recent experimental study reported a $27\%$ drag reduction in a high-speed towing tank \citep{xu2021superhydrophobic}.  
  In channel flow,
Stancanelli $\textit{et al.,}$  \cite{stancanelli2024magnetic} used a wall-attached magnetic fluid film to reduce wall drag by up to $90 \%$. They found that the strong damping of wall friction appears from the co-existence of slip and waviness at the coating interface, and the latter is the key to complete wall drag reduction across laminar and turbulent flow regimes. 
 
  Ideally, if turbulence completely collapsed, 
  great energy saving would be achieved. The findings of Hof $\textit{et al.,}$ \cite{hof2010eliminating}
  show that this strategy is actually
  possible, and it has been achieved for a series of Reynolds numbers through radial jets, axial injection, moving walls, rotors, and in numerical experiment through volume forcing, all of which aim to flatten the laminar velocity profile \citep{kuhnen2018destabilizing}. The TG of the flattened laminar velocity profile, which plays an essential role in sustaining shear turbulence with linearly stable laminar flow \citep{kim2000linear}, is found to be greatly reduced. SSP is consequently disrupted when the TG is suppressed. 

 The strategy of laminarising turbulence is promising, possible in
 principle, but it remains an important open problem 
 how to efficiently achieve it.
 While there are many possibilities, here, we restrict ourselves to body forces, which could be considered as a generalised proxy for the  strategies mentioned earlier, but perhaps more importantly at this stage, body forces are  amenable to the  
 fully nonlinear variational optimisation approach.
The core of this technique is to optimize an objective function, constrained by the fact that each disturbance evolves subject to the full Navier-stokes equations \citep{kerswell2018nonlinear}. Flexibility in the method allows us to select the form of the desired target by designing suitable objective functions. Over the past ten years, this method has been successfully applied to find the minimal seeds for transition to turbulence \citep{pringle2010using, pringle2012minimal,pringle2015fully}, for finding periodic orbits \citep{parker2022variational} and heteroclinic orbits \citep{farano2019computing} in shear flows, and to enhance the nonlinear stability of the laminar flow \citep{rabin2014designing, marensi2019stabilisation}.   
  
 Recently, Maresnsi $\textit{et al.,}$ \cite{marensi2020designing} considered a purely damping body force to control turbulence.
 This is considered to be a `passive' control strategy, as it does not inject energy into 
 the flow at any point.
 The variational approach was employed to optimise over the space of forces of the form $ \vec{F}=-\chi(\vec{x}) \vec{u}_{tot}$, 
 where $\chi$ is positive and 
 corresponds to the inverse of the permeability of a porous medium -- the scalar field $\chi$ could be considered as a proxy for a permeable baffle introduced into the flow.
It was 
found that laminarisation can be realised by a baffle which damps the flow more strongly near the wall than in the centre, by targeting the region of turbulent energy production. This discovery was quite different from the results of K\"hnen $\textit{et al.}$ \cite{kuhnen2018destabilizing}, who found that accelerating the flow near the wall can eliminate turbulence. However, the mechanism behind the two methods may be quite different. 
The former employs a time-dependent passive damping body force, 
while the
latter applied a time-independent `active' force, which inputs energy to the flow near the wall and reduces the energy of the flow in the centre.  
Ding $\textit{et al.,}$ \cite{ding2020stabilising}  maximised the energy stability of a baffle-modified laminar flow using a `spectral' approach.
It is found that the optimal baffle is always axisymmetric and streamwise independent, retaining just radial dependence. Numerical simulations show that the optimal baffle can laminarise turbulence efficiently at moderate Reynolds numbers $(Re\leq 3500)$, and an energy-saving regime can be also identified.
While the passive `baffle' force of Marensi $\textit{et al.,}$ \cite{marensi2020designing} might be easier to implement in practice, it may be
less efficient than the active body force of K\"uhnen $\textit{et al.,}$ \cite{kuhnen2018destabilizing}, due to the additional baffle drag.
As the active force has more freedom to shape the flow, there is potential for higher net-power saving. 
    
  In this paper, we use the Lagrangian optimisation technique to explore the most efficient active body force to laminarise turbulence. The plan of the paper is as follows. In \S 2, we present direct numerical simulation (DNS) of pipe flow, the variational equations and optimisation results at $Re=2400$ and $Re=3000$. In \S 3, based on the results of the nonlinear optimisation, we propose a linear optimisation to find the optimal forced laminar profile with minimal TG, and explore whether the reduction of TG alone can eliminate turbulence. This approach has more in common with Ding $\textit{et al.,}$ \cite{ding2020stabilising}, but is focused on reducing linear transient amplification of disturbances, rather than maximising the energy stability of the forced laminar flow. Finally, we discuss the mechanism of laminarisation, and comment on the general philosophy of the linear versus nonlinear optimisation approaches.
\section{Nonlinear optimisation}
  \label{sec:headings}

In this section, we adapt the approach developed by Marensi $\textit{et al.,}$ \cite{marensi2020designing} to produce laminarisation by a passive force to the 
case of an active force.
 
  \subsection {Formulation}
  \subsubsection {Simulations and body forces}
 
   We consider the flow of an incompressible Newtonian fluid through a straight cylindrical pipe of 
   radius $R$ and length $L$. 
   The flow has a constant mass flux with bulk speed $U_b$ and is driven by an externally applied pressure gradient. The flow is described using cylindrical coordinates $\{r, \theta, z\}$, where $z$ is aligned with the pipe axis. Length scales are non-dimensionalised by the pipe radius $R$ and velocity components by the unforced laminar centreline velocity $2U_b$. 
   The Reynolds number is $Re= 2U_{b}R/\nu$, where $\nu$ is the kinematic viscosity of the fluid. 
   Hereafter we use dimensionless units.

    Calculations are carried out with the open-source code openpipeflow.org \citep{willis2017openpipeflow}. At each time step, the unknown variables, i.e. the velocity and pressure perturbations $\{\vec{u}, p\}$ are discretised in the domain $\left\{ r, \theta, z \right\}=[0,1]\times[0, 2\pi)\times[0, 2\pi/k_0)$, where $k_0=2\pi/L$, using Fourier decomposition in the azimuthal and streamwise direction, and finite difference in the  radial direction 
    \begin{equation}
        \{\vec{u}, p\}(r_s, \theta, z)=\sum_{k<|K|} \sum_{m<|M|}\{\vec{u}, p\}_{skm}e^{i(k_0kz+m\theta)}
    \end{equation}
    where $s=1,...,S$ and the radial points are clustered close to the wall. Temporal discretisation is via a second-order predictor-corrector scheme, with an Euler predictor for the nonlinear terms and the Crank-Nicolson corrector.  The Reynolds numbers $Re=2400$ and $Re=3000$  are adopted, and a domain of length $L=10$, 
    to maintain manageable computation requirements. We use a resolution of $S=64, M=32, K=36$ at $Re=2400$ and  $S=64, M=42, K=48$ at $Re=3000$, and the size of the time step is $\Delta t=0.01$. 
    
   We consider two forms of body force:
  \begin{equation}
  \label{eq:F1F2}
  \vec{F}_1=F_z(r)\hat{\vec{z}},
  \qquad 
  \vec{F}_2=F_r(r,\theta)\hat{\vec{r}}+F_\theta(r,\theta)\hat{\vec{\theta}}+F_z(r,\theta)\hat{\vec{z}}. 
  \end{equation}
  We denote  the resulting laminar forced flows by $\vec{u}_{lam}$.
  In the iterative optimisation method, to obtain the specific form of body force, a filter $\mathcal{F}$ is applied to remove unwanted Fourier modes ($k\ne 0$ and/or $m\ne 0$) and components of the force from the initial guess, if necessary, and from subsequent updates to the force.

  Using cylindrical coordinates
  $\vec{x}=(r, \theta, z)$,
  total velocity and pressure fields are decomposed as $\vec{u}_{tot}=U(r)\vec{\hat{z}}+\vec{u}(\vec{x},t)$ and $p_{tot}=P(z)+p(\vec{x}, t)$, where $\vec{u}=(u_r, u_{\theta}, u_z )$  and $p$ are the deviations from the unforced laminar velocity $U(r)=(1-r^2)$ and pressure field $P=-\frac{4z}{Re}$, respectively.  The problem is governed by the Navier--Stokes and continuity equations for incompressible Newtonian fluid flows: 
  \begin{equation}
       	\frac{\partial \vec{u}}{\partial t}+U\frac{\partial \vec{u}}{\partial z}+u_{r}{U}'\hat{\vec{z}}-\vec{u}\times\vec{\nabla}  \times \vec{u} +\vec{\nabla} p 
        + \frac{4\kappa}{Re}\hat{\vec{z}}
        - \frac{1}{Re}\vec{\nabla} ^{2}\vec{u}- 
        \vec{F}=0
   \end{equation}
   \begin{equation}
       	\vec{\nabla}\boldsymbol{\cdot}\vec{u}=0,
   \end{equation} 
    where $p$ and $\vec{u}$
    are assumed to be spatially periodic
    over a length $L$.  The scalar
    $\kappa(t)$ measures the relative 
    additional pressure
    gradient required to maintain the fixed
    flow rate 
relative to the laminar case,
    \begin{equation}
     1+\kappa=\frac{\left<\partial p_{tot}/\partial z \right>}{\left<\partial P/\partial z \right>},
     \label{extrap}
    \end{equation}
    where the angle brackets $\left< \bullet \right>$ indicate the volume integral:
    \begin{equation}
     \left< (\bullet )\right>=\int_{0}^{L}\int_{0}^{2\pi}\int_{0}^{1}(\bullet )\,r\,\mathrm{d}r\,\mathrm{d}\theta \, \mathrm{d}z.
    \end{equation}
    Periodic boundary conditions are adopted in the streamwise and azimuthal direction, and no-slip conditions are imposed  on the walls.
    The streamwise, azimuthal and cylindrical-surface averages are introduced as follows 
    \begin{equation}
     \overline{(\bullet )}^{z}=\frac{1}{L}\int_{0}^{L}(\bullet )\,\mathrm{d}z,  
     \;\;\;
     \overline{(\bullet )}^{\theta}=\frac{1}{2\pi}\int_{0}^{2\pi}(\bullet )\,\mathrm{d}\theta, 
     \;\;\;
     \overline{(\bullet )}^{\theta,z}=\frac{1}{2\pi L}\int_{0}^{L}\int_{0}^{2\pi}(\bullet)\mathrm{d}\theta \, \mathrm{d}z.
    \end{equation}   
     From the average force balance in the streamwise direction, it can be determined that
     \begin{equation}
      \kappa(t)=-\left.\frac{1}{2}\frac{\partial \overline{u_z}^{\theta,z}}{\partial r}\right|_{r=1}+\frac{Re}{2}\int_{0}^{1}\overline{-\vec{F}\boldsymbol{\cdot} \hat{\vec{z}}}^{\theta,z}r\,\mathrm{d}r,
      \end{equation}
    where the first term on the right hand is the compensation for loss of flux due to the frictional drag, and the second term is the flux loss/gain due to the body force.

\subsubsection{Variational optimisation}

    To construct Lagrange equations for finding a body force that laminarises turbulence, we need to pick a suitable physical quantity for the objective function that describes the decay of turbulence. Referring to the nonlinear optimisations of Monokrousos $\textit{et al.,}$\cite{monokrousos2011nonequilibrium} and Marensi $\textit{et al.}$ \cite{marensi2020designing}, we first consider the total viscous dissipation ${\mathcal D}_{tot}$, which is proportional to the frictional drag by the flow, as follows:
    \begin{equation}
     \mathcal J_{1}= \sum_{n=1}^{N}\frac{1}{T}\int_{0}^{T}{
     {\mathcal D}}_{tot, n}{(t)} \mathrm{d}t= \sum_{n=1}^{N}\frac{1}{T}\int_{0}^{T}\frac{1}{Re} \left \langle
     (\vec{\nabla} \times \vec{u}_{tot, n}(\vec{x},t))^{2} \right \rangle \mathrm{d}t
    \end{equation} 
    To increase the robustness of the obtained optimal body force, we consider $N > 1$ initial conditions (ICs) for the turbulent velocity field.  Typically $N=10$ is found to be practical and sufficient to determine the fundamental properties of the optimal forcings. Here, we minimize the time-averaged objective function to smooth the hypersurface of the Lagrangian, improving convergence properties \citep{monokrousos2011nonequilibrium,marensi2020designing}. The target time $T$ needs to be sufficiently long to ensure full laminarisation, but large $T$ is computationally expensive,
    not just due to longer integration times, 
    but due to its effect on the accuracy of the 
    computed gradient.
    Here, $T=300$ is adopted in the optimisations,
    which is approximately twice the time 
    over which peak linear growth can be obtained
    at $Re=2400$.
    In addition to ${\mathcal D}_{tot} $, we consider a second simpler and more direct objective function to describe the decay of turbulence,  $E_{3d}=E_{tot}-E_{k=0}$, i.e.\ the energy of the streamwise-dependent component of the flow.   This component is known to 
    decay much more rapidly upon laminarisation
    than the full energy, and has been used
    when measuring the lifetime of puffs 
    \citep{avila2010transient}. 
    Here the objective function is
    \begin{equation}
    \label{eq:J2}
    \mathcal J_{2}= 
    \frac{1}{T}\int_0^T \sum_{n=1}^N E_{3d,n}(t) \mathrm{d}t = 
    \frac{1}{T}\int_{0}^{T}\sum_{n=1}^{N}\left \langle \frac{1}{2}\mathcal F_{u}(\vec{u}_{n}(\vec{x},t))^2 \right \rangle\mathrm{d}t,
    \end{equation} 
    where the $\mathcal F_{u}(u)$ is a filter that removes the streamwise-dependent Fourier modes of the argument. We note that we have not tried the 
    the total energy input as an objective
    function, suggested by Marensi $\textit{et al.}$ \cite{marensi2020designing}.
    There, the force was purely damping,
    i.e.\ $\vec{F}\cdot\vec{u}$ was negative everywhere.  Here the product can be
    of both signs, which could lead to cancellation in the integral, and hence potential ambiguity about how the energy
    should be calculated.  Further comment
    is made in Appendix A.
    
    The efficiency of the variational model depends substantially on the choice of the objective function, as it determines the form of  particular terms in the adjoint equations.  We will hence compare $\mathcal{J}_1$ and $\mathcal{J}_2$ as objective functions and their corresponding optimal forcing.  These functions are minimised with the constraints of the three-dimensional Navier-Stokes equation, constant mass and a given amplitude of the body force. The Lagrangian equation is 
    \begin{eqnarray}
     \mathcal L=&&\mathcal J +\lambda( \frac{1}{2}\langle\vec{F}^{2}\rangle-A_{0})+\sum_{n=1}^{N}\int_{0}^{T}\left \langle \vec{v}_n\boldsymbol{\cdot}(\vec{NS}(\vec{u}_n)-\vec{F})\right \rangle\mathrm{d}t\nonumber\\ 
     && +\sum_{n=1}^{N}\int_{0}^{T}\left \langle \Pi_n (\vec{\nabla}\boldsymbol{\cdot} \vec{u}_n) \right \rangle\mathrm{d}t+ \sum_{n=1}^{N}\int_{0}^{T} \Gamma_n(t)\left \langle (\vec{u}_n\boldsymbol{\cdot} \vec{\hat{z}}) \right \rangle\mathrm{d}t
    \end{eqnarray}
    Here, we use the $L_2$ norm to measure the amplitude of the body force, rather than the $L_1$ norm used in Marensi $\textit{et al.}$ \cite{marensi2020designing}, as our force may be locally negative.  $\lambda, \vec{v}_n, \Pi_n,\Gamma_n$ are Lagrange multipliers. Taking variations of $\mathcal L$ 
    with respect to each variable
    and setting them equal to zero, we obtain the following set of Euler-Lagrange equations for each turbulent field. In the  current case, the different objective functions only make a different adjoint equation for $\vec{v}_n$, while other equations are all the same.  When ${\mathcal D}_{tot}$ is the objective function, the adjoint equations are 
    \begin{eqnarray}
    \label{adjoint-D}
    \frac{\partial \mathcal L} {\partial \vec{u}_n}=&&\frac{\partial \vec{v}_n}{\partial t}+U\frac{\partial \vec{v}_n}{\partial z}-v_{z,n}{U}'\vec{\hat{r}}+\vec{\nabla}\times(\vec{v}_n\times \vec{u}_n)- \vec{v}_n\times\vec{\nabla} \times \vec{u}_n \nonumber\\
    &&+\vec{\nabla} \Pi +  \frac{1}{Re}\vec{\nabla} ^{2}\vec{v}_n-\Gamma_n\vec{\hat{z}}+\frac{2}{Re}\vec{\nabla}^{2}\vec{u}_{tot,n}=
    {\bf 0}.
    \end{eqnarray}
     When $E_{3d}$ is the objective functional, the adjoint equations are
    \begin{eqnarray}
    \label{adjoint-E}
    \frac{\partial \mathcal L} {\partial \vec{u}_n}=&&\frac{\partial \vec{v}_n}{\partial t}+U\frac{\partial \vec{v}_n}{\partial z}-v_{z,n}{U}'\vec{\hat{r}}+\vec{\nabla}\times(\vec{v}_n\times \vec{u}_n)- \vec{v}_n\times\vec{\nabla} \times \vec{u}_n \nonumber\\
    &&+\vec{\nabla}\Pi +  \frac{1}{Re}\vec{\nabla} ^{2}\vec{v}_n-\Gamma_n\vec{\hat{z}}+\mathcal F_{u}(\vec{u}_{n}(x,t))={\bf 0}.
    \end{eqnarray}
The adjoint variable satisfies the
 continuity equation
    \begin{equation}
    \frac{\delta \mathcal L}{\delta p_n}=\vec{\nabla} \boldsymbol{\cdot} \vec{v}_n =0,
    \end{equation}
    and the compatibility condition is given by
    \begin{equation}
    \frac{\delta \mathcal L}{\delta \vec{u}_n(x,T)}=\vec{v}_n(\vec{x},T)=0.
    \end{equation} 
    The optimality condition is
    \begin{equation}
    \frac{\delta \mathcal L}{\delta \vec{F}}=\lambda \vec{F} +\vec{\sigma}=0 \, ,
    \end{equation}
    where \begin{equation}
    \vec{\sigma}(r,\theta, z)=\sum_{n=1}^{N}\int_{0}^{T}\mathcal{F}(\vec{v}_{n})dt
    \end{equation}
    is a vector function of position. 
    The filter $\mathcal{F}$, introduced
    at (\ref{eq:J2}), removes unwanted spatial
    dependence from the update to $\vec{F}$.
    
    As the optimality condition is not satisfied automatically, $\vec{F}$ is moved in the direction of reduced $\mathcal L$, to achieve a minimum where $ \delta\mathcal L/\delta \vec{F} $ should vanish. The minimisation problem is solved numerically using an iterative algorithm similar to that applied in Pringle $\textit{et al.}$ \cite{pringle2012minimal}. The update for $\vec{F}$ at $(j+1)\mathrm{th}$ iteration is
    \begin{equation}
    \vec{F}^{(j+1)}=\vec{F}^{(j)}-\epsilon\frac{\delta \mathcal L}{\delta \vec{F}^{(j)}}
    \end{equation}
    The strategy used to find $\lambda$ and the control of $\epsilon$ is that of Pringle $\textit{et al.}$ \cite{pringle2012minimal}. 

    A key observation is that, 
    for the minimal-seed problem, the variational method breaks
    down when the initial amplitude of the 
    velocity perturbation is sufficiently large
    that the flow reaches turbulence.  
  Marensi $\textit{et al.,}$ \cite{marensi2020designing} therefore proposed that 
    one should start with a sufficiently large $A_0$ that all $N$ turbulent ICs are
    laminarised.  Then $A_0$ is decreased gradually until it reaches a critical value $A_{crit}$ at which the flow remains turbulent for one or more ICs, i.e.\ not all cases can be laminarised.
    
\subsection{Results for nonlinear optimisation}
  \subsubsection{Preliminary calculations}
\begin{figure}
    \centering
    \includegraphics[angle=0,width=1\textwidth]{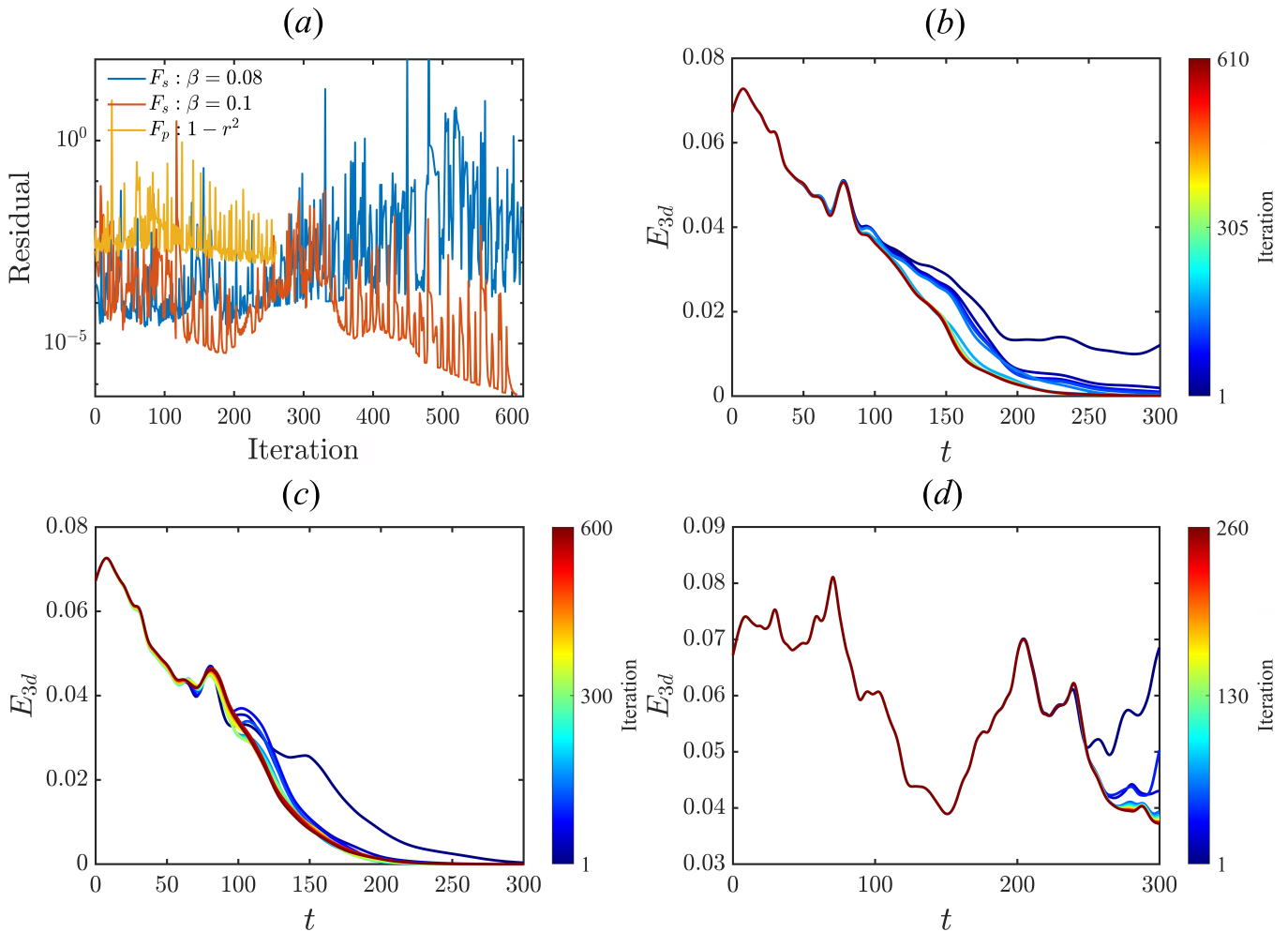}
    \caption{$(a)$ The residuals ($\delta \mathcal{L}/\delta {\vec F_1}$) of the preliminary cases ($T=300, Re=2400, N=1, \mathcal{J}=\mathcal{J}_1$ based on $\mathcal D_{tot}$) for three different initial $\vec{F}^{(0)}$. $(b-d)$ $E_{3d}(t)$ during the optimisation process 
    for $\vec F_s:\beta=0.08$, $\vec F_s:\beta=0.1$ and $\vec F_p$ respectively.  
    }
    \label{TestoneIC}
    \end{figure}

    We start with $\mathcal{J}_{1}$, based on the total dissipation $\mathcal{D}_{tot}$, as the objective function and test the algorithm 
    with one IC ($N=1$) and $T=300$ to find the optimal body force of form $\vec{F}_1$ at $Re=2400$.  We compare three 
    initial guesses $\vec{F}^{(0)}$ for the body force.  The first two use the body force of Hof $\textit{et al.,}$ \cite{hof2010eliminating} and Song $\textit{et al.}$ \cite{song2014direct} at two different amplitudes.  
    We mark this type of starting body force $\vec{F}^{(0)}=\vec{F}_s$. 
    This force is purely axial and has been shown to laminarise turbulence up to quite high Reynolds numbers. The corresponding laminar forced flow $\vec{u}_{lam}$ is parameterised by the centreline deviation $\beta$  from the laminar parabolic profile and is given by
    \begin{equation}
        \label{songprofile}
       	\vec{u}_{lam}(r;\beta)=(1-\beta)\left(1-\frac{\cosh(cr)-1}{\cosh(c)-1}\right)\hat{\vec{z}},
    \end{equation} 
    where the value for $c$ is determined by the condition that the dimensionless
    mean flux satisfies $U_b=1/2$ for a given $\beta$.  We start with the cases 
    $\beta=0.1, c=2.15$ and $\beta=0.08, c=1.87$.
    Given the flattened laminar velocity profile \eqref{songprofile}, the
    corresponding force can be inverted from the Navier--Stokes equation for laminar flow. 
    The corresponding force amplitudes are $A_0=8.67e-6$ and $A_0=5.28e-6$, respectively. 
    To examine how the method copes with
    an initial force that does not suppress
    turbulence, we consider a third intial
    force that is parabolic,
    $\vec{F}^{(0)}=\vec{F}_p=\gamma_1(1-r^2)\hat{\vec{z}}$,
    where $\gamma_1$ is a positive constant. 
    This force has the opposite effect on the flow to $\vec{F}_s$, making the mean streamwise velocity profile more parabolic and less plug-like. 
    The amplitude $\vec{F}_p$ is scaled by selecting $\gamma_1$ so that $A_0=8.67e-6$, for comparison with the other initial forces. Figure \ref{TestoneIC}$(a)$ shows the residuals of the optimisations starting with these three initial body forces. It is found that the algorithm converges well only for the initial guess $\vec{F}_s$ with $\beta=0.1$, while for the other two cases, a clear drop of the residual is not observed.  
    Figure \ref{TestoneIC}$(b-d)$ show the curves of $E_{3d}(t)$ 
    starting from these three different $\vec F^{(0)}$.
    (Although the objective function to minimise is based on $\mathcal{D}_{tot}$, turbulence suppression is more clearly evident by observing $E_{3d}$, as discussed later
    in section 2.2.2.)
    For the case $\vec F^{(0)}=\vec F_s$ with $\beta=0.08$, we see that the turbulence decays, but is not fully laminarised by the initial force (darkest blue line corresponds to the first iteration).  Only after many iterations is the flow laminarised, but the residual has not yet dropped significantly. 
    This suggests that achieving the final optimal force may still require many more iterations.  For the case with $\beta=0.1$,  the flow is almost laminarised from the beginning, i.e.\ for the first iteration, and the optimised forcing achieves earlier laminarisation. For the case with $\vec{F}^{(0)}=\vec{F}_p$, 
    the dynamics remains turbulent, 
    the optimisation struggles, only 
    managing to reduce the energy near the end of the time series.
    These comparisons stress the importance of selecting an initial force that laminarises the flow,  before optimising that force.
    

\begin{figure}
     \centerline{\includegraphics[angle=0,width=1\textwidth]{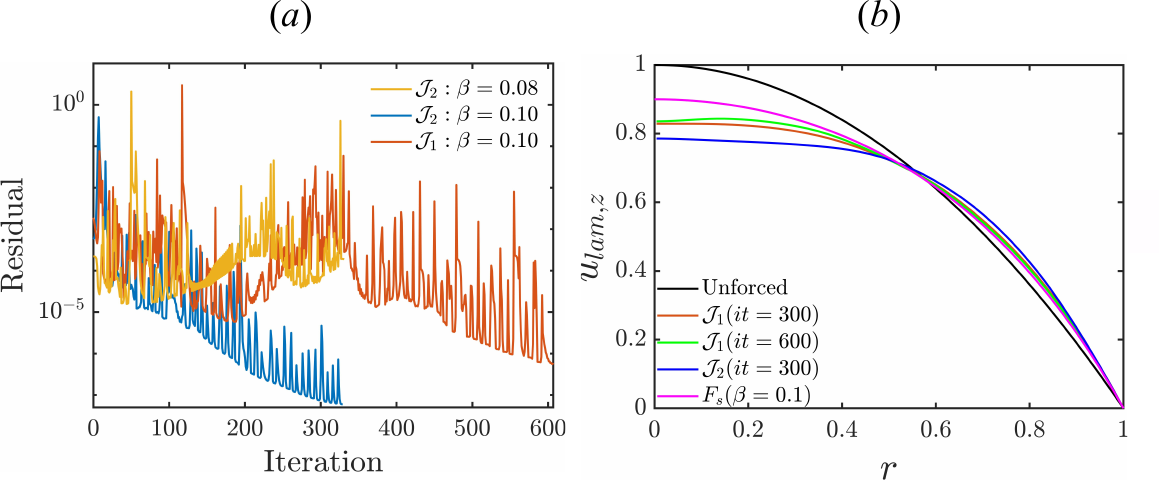}}  
       \caption{$(a)$ The residuals ($\delta \mathcal{L}/\delta {\vec F_1}$) of the preliminary cases ($T=300, Re=2400, N=1, \mathcal J= \mathcal J_2$ based on $E_{3d}$) for initial guesses $\vec{F}^{(0)}=\vec{F}_s$ with $\beta=0.08$ and $\beta=0.1$, compared with case of $\mathcal J_1$ with  $\vec{F}^{(0)}=\vec{F}_s(\beta=0.1)$.  
       $(b)$ The forced laminar velocity profile for the forcing optimised by two types of objective functions, where $it=300$ means the profile is picked in the 300th iteration. The magenta line corresponds to the input forced profile (i.e. Song's \citep{song2014direct} laminar velocity profile  given by \eqref{songprofile} with $\beta=0.1$). 
       }
    \label{E3dN1}
    \end{figure}

    The initial body forces $\vec{F}_s$ with $\beta=0.08$ and $\beta=0.1$ were also optimised with $\mathcal{J}_2$, based on $E_{3d}$, as the objective function. Residuals are shown in figure \ref{E3dN1}$(a)$. The residual for the case $\vec{F}^{(0)}=\vec F_s (\beta=0.08)$ still struggles to reduce, due to the only partial laminarisation, 
    but for the case with $\vec{F}^{(0)}=\vec F_s (\beta=0.1)$ the residual decreases more straightforwardly and rapidly than with objective function $\mathcal{J}_1$. 
    Figure \ref{E3dN1}$(b)$ shows the effect of the optimisation on the forced laminar profiles $u_{lam,z}(r)=\vec{u}_{lam}\boldsymbol{\cdot} \hat{\vec{z}}$ obtained with the two objective functions for $\vec{F}^{(0)}=\vec F_s (\beta=0.1)$.  Compared with the input forced profile (Song's profile \citep{song2014direct} given by \eqref{songprofile}, see magenta profile in  \ref{E3dN1}$(b)$),  
    the optimised profiles obtained for $\mathcal J_1$ 
    and $\mathcal{J}_2$ are slightly
    more flattened, but note that they
    are not quite the same.  The reason for 
    this difference is discussed in the 
    next subsection.
    
    A factor that may affect the 
    rate of convergence  
    is that, $E_{3d}$ is smoother than $\mathcal{D}_{tot}$, and the adjoint equation is 
    forced by a spatially smoother term, proportional to filtered $\vec{u}$ (\ref{adjoint-E}) versus $\boldsymbol{\nabla}\times\vec{u}$ (\ref{adjoint-D}).
    This is expected to lead to 
    better convergence properties
    for the $\mathcal{J}_2$ case, improving 
    the rate of convergence and potentially reducing the chances of getting stuck in local minima or saddle points of the Lagrangian hypersurface. 


\subsubsection{Minimal amplitude $A_0$ for $\mathcal{J}_1$ and $\mathcal{J}_2$}

    To improve the robustness of the final optimal force, we now increase to $N=10$ ICs.  For more ICs, a larger $\beta$ is required to ensure that all ICs are laminarised by the initial force.  We take $\vec{F}^{(0)}=\vec{F}_s$ with $\beta=0.2, c=3.5935$, having a corresponding amplitude $A_0=4.82e-5$ at $Re=2400$.  At $Re=3000$, we start with $\beta=0.25, c=4.4912$ and corresponding amplitude $A_0=6.12e-5$.

\begin{figure}     \centerline{\includegraphics[angle=0,width=1\textwidth]{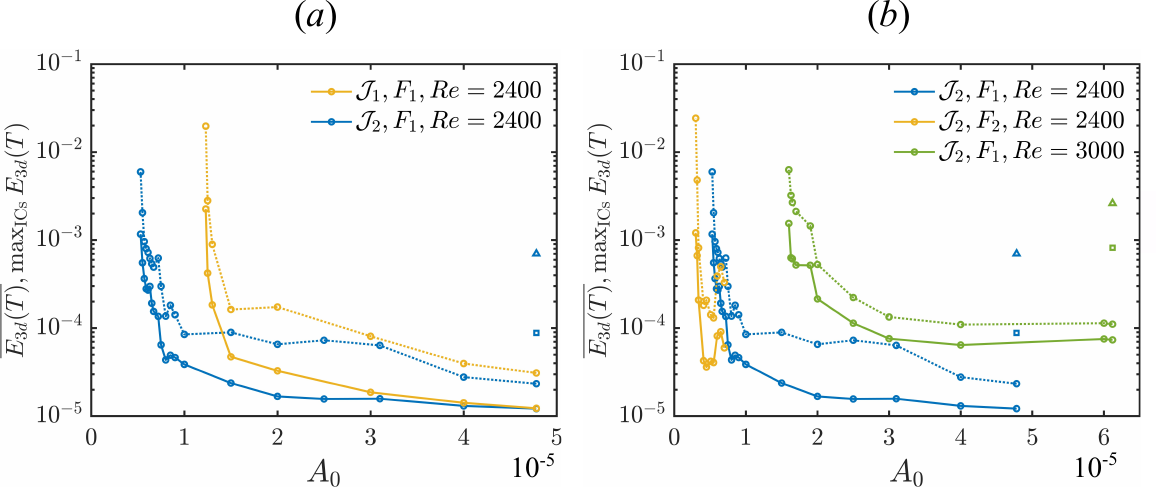}}  
       \caption{ 
       $\overline{E_{3d}(T)}$ (average over the $10$ ICs) (solid line) and maximal ${E_{3d}(T)}$ of $N=10$ ICs (dashed line) as a function of force amplitude. (\textit{a}) 
       Comparison of objective function at $Re=2400$, force form $\vec{F}_1$.
       (\textit{b}) Comparison of form of force, Reynolds number and objective function $\mathcal{J}_2$. The triangle and square represent $\max_\mathrm{ICs}E_{3d}(T)$
       and $\overline{E_{3d}(T)}$ of the input unoptimised forced flow.
    }
    \label{OFcomparison}
    \end{figure}

    We now compare the
    minimal force of $\vec{F}_1$ for the two objective functions. 
    The ensemble average of ${E_{3d}(T)}$, denoted by $\overline{E_{3d}(T)}$,
    is used to measure the final laminarisaion
    over the $N$ cases,
    and the maximum $E_{3d}(T)$, marked as $\max_\mathrm{ICs}E_{3d}(T)$, more 
    sensitively indicates whether all cases
    have been laminarised. 
    Starting from the amplitude of the input forcing (isolate triangle and square in  figure \ref{OFcomparison}), we gradually reduce the force amplitude $A_0$.
    It should be noted that $A_0$ cannot be reduced by too large a step, otherwise the initially unoptimised force may not laminarise all ICs, causing the algorithm to struggle to converge.
%
    It is observed that $\overline{E_{3d}(T)}$ and $\max_\mathrm{ICs}E_{3d}(T)$ reduces substantially with the first optimisation, meaning an earlier laminarisation is achieved.  
    Then,  $\overline{E_{3d}(T)}$ and $\max_\mathrm{ICs}E_{3d}(T)$ gradually increase with the reduction of force amplitude, then jump up at a critical force amplitude, $A_{crit}$,
    which gives the minimal force for laminarisation. 
    An earlier jump can be seen in
    $\max_\mathrm{ICs}E_{3d}(T)$,
    but at quite close $A_0$ to for the jump in $\overline{E_{3d}(T)}$.
    In the limit $N\to\infty$
    the non-laminarisation of only one
    case would give no detectable jump, but large $T$ may give sufficient time for trajectories starting from all ICs (or at least a significant proportion) to 
    reach the laminar state.
    (In the limit $T\to\infty$, reduction of  $A_0$ would produce a step function, as discussed in Kerswell $\textit{et al.}$ \cite{kerswell2018nonlinear}, while a finite $T$ produces a smoothed version.  Here, we see that $T=300$ is sufficiently large to locate the jump.)
   
    As shown in figure \ref{OFcomparison}(\textit{a}), from the comparison of the results for $\mathcal{J}_1$ and $\mathcal{J}_2$, the latter appears to be more efficient, as it can achieve a smaller minimal force, yet the form of the force appears to be almost the same (not shown). Given that both objective functions are intended to measure the presence of turbulence versus laminarisation, it is not immediately obvious that there should be a difference.  It is interesting to inspect why.

    As $\mathcal{D}_{tot}$ measures the friction drag, $\mathcal{J}_1$ might be considered a better choice.
    The total dissipation can be divided into two parts as
    \begin{equation}
    \mathcal D_{tot}=\frac{1}{Re}\vec{\nabla}\times \vec{u}_{tot}=\frac{1}{Re}\vec{\nabla}\times (\vec u_{lam}+\vec{u})=\mathcal D_{lam}+\mathcal D_{tur},
    \end{equation}
    where $\vec u_{lam}$ is the forced laminar 
    flow and $\vec{u}$ is the turbulent perturbation from laminar solution. It is expected that the optimisation reduces $\mathcal D_{tur}$ to zero, and that $\mathcal D_{lam}$ is reduced to a minimum. This places two constraints on the algorithm.  In other words, optimisation with $\mathcal{J}_1$ must work on laminarisation with the least dissipation of the forced laminar profile simultaneously. These two targets push the laminar velocity profile
    in opposing directions -- the target of laminarisation favors a flattened forced laminar velocity profile, 
    but this raises the gradient near the wall
    and therefore is counter to minimal
    dissipation (see figure \ref{E3dN1}($b$)).
    On the other hand, $E_{3d}$ only measures the strength of turbulence, thus the focus for $\mathcal{J}_2$ is purely on eliminating turbulence.  
    
    Although optimisation with $\mathcal{J}_2$ may lead to a larger laminar dissipation, convergence is found to be easier, requiring far fewer iterations, and a smaller minimal force can be obtained. Therefore, $\mathcal{J}_2$  is adopted as the objective function throughout the rest of this work.

\subsubsection{Form of the minimal force}
    The body forces of forms $\vec{F}_1$ and $\vec{F}_2$ of (\ref{eq:F1F2})
    are optimised at $Re=2400$, and $\vec{F}_1$ is also optimised at $Re=3000$.  As $\vec{F}_2$ is a less constrained form than $\vec{F}_1$, the minimal force $\vec{F}_1$ is used as an initial guess for $\vec{F}_2$, with a small perturbation that breaks the rotational symmetry and that includes azimuthal and radial components. By reducing the force amplitude $A_0$ gradually, we obtained  $\overline{E_{3d}(T)}$ and $\max_\mathrm{ICs}E_{3d}(T)$as a function of $A_0$, shown in figure \ref{OFcomparison}(\textit{b}). Clear jumps can be observed for both Reynolds numbers.
    A smaller force amplitude for $\vec{F}_2$ is achieved, as 
    might be expected, since it is less constrained, although the difference is small.
    Compared to $\vec{F}_1$, the small reduction of force amplitude for $\vec{F}_2$ indicates that the cross-stream components of body force do not play a significant role in eliminating turbulence. This is quite different from the case of Marensi $\textit{et al.}$ \cite{marensi2020designing}, in which the cross-flow components are necessary for laminarisation. It may be attributed to the time-independent minimal force used in our optimisation, which therefore cannot oppose at all times the vortices that meander.

\begin{figure}
       \centering
      \includegraphics[angle=0,width=1\textwidth]{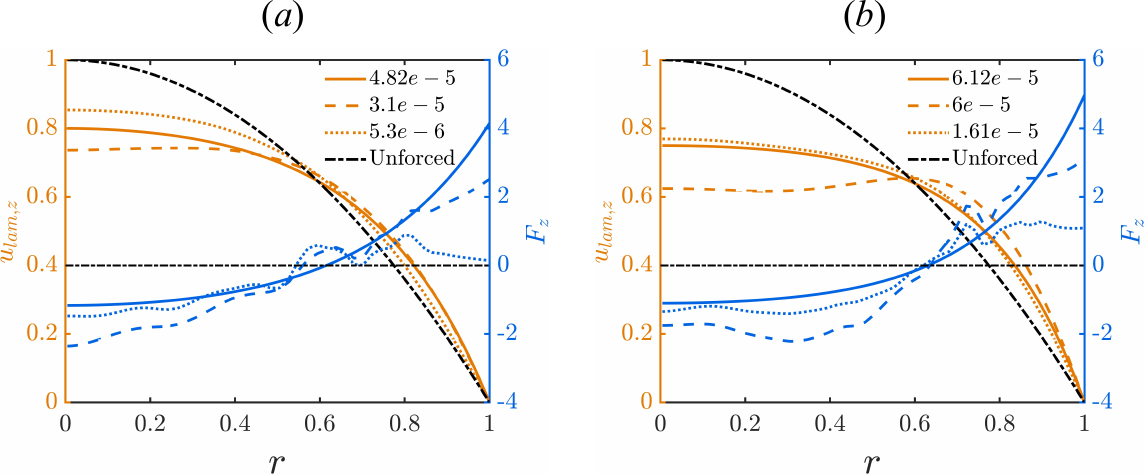}  
       \caption{Profile of  the optimised  body force $\vec F_1$,
       and forced laminar profile at ($a$) $Re=2400$ and  ($b$) $Re=3000$.  The three force amplitudes represent the input force (solid line), first optimisation (dashed) and the minimal force (dotted), respectively. The laminar unforced (parabolic) profile and corresponding zero forcing are shown with dash-dotted black line. 
       }
       \label{forceandUz}
    \end{figure}

    The profile of the initial input force $\vec F^{(0)}$, the optimised $\vec F_1$  and their corresponding laminar solutions at $Re=2400$ and $Re=3000$ are shown in figure \ref{forceandUz}. 
    Although the force profile is not smooth, which may be due to the limited number of ICs used, it is clear that the resulting forced laminar profile itself is smooth.  
    As a result, the force laminar velocity profile is flattened.  At $Re=2400$, comparing the initial 
    and optimised body force,
    the first optimisation brings stronger damping in the centre of pipe, leading to a more flattened forced laminar profile. The force near the wall is reduced. This redistribution of force indicates that damping of flow in the centre of pipe, rather than the acceleration of flow close to the wall, is the priority for laminarisation. For the minimal force, the positive force near the wall becomes very small. 
    A similar conclusion can be drawn at $Re=3000$-- the initial optimisation
    flattens the profile, but notice that 
    the optimal profile at the minimal 
    amplitude is close to the initial
    profile (however the latter requires a larger amplitude
    to produce!).
    Comparing the optimals at 
    different $Re$, a more flattened forced laminar profile is required for laminarisation at higher Reynolds number.
    
    Starting with other initial guesses, the optimisation did not find any other 
    body force, it is fundamentally still of the same form as that of K\"uhnen $\textit{et al.}$ \cite{kuhnen2018destabilizing}, indicating that the body force which flattens the laminar forced profile is very efficient for eliminating turbulence.  
    However, we add a cautionary note that 
    we have been unable 
    to try many initial guesses, due to the 
    requirement discussed in section 2.2.1 that any starting force must laminarise the flow even before optimisation.
       
\begin{figure}
       \centering
       \includegraphics[angle=0,width=0.9\textwidth]{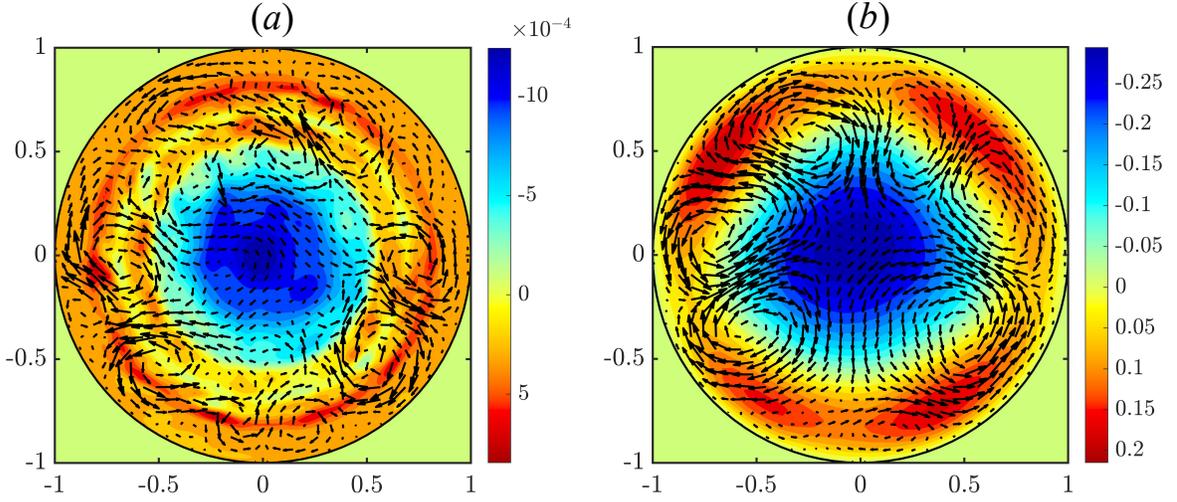}  
       \caption{ (\textit{a}) Contour of streamwise component of 
       and arrows for the cross-stream components of optimised $\vec F_2$. 
       the amplitude of streamwise component is $O(10^{-5})$, while the amplitude of cross components are $O(10^{-7})$.
       (\textit{b}) Ensemble-averaged velocity $u$ of the 10 ICs at $Re=2400$. 
       }
       \label{F3Umean}
    \end{figure}

    The body force $\vec{F_2}$ is shown in figure \ref{F3Umean}(\textit{a}), which suggests complicated roll structures, while the streamwise component still damps the flow in the centre and accelerates the flow near the wall. However, the magnitude of the cross-stream components is low, and if we compare the cross-stream components of the force with the ensemble-averaged velocity field of the $N=10$ chosen ICs, it is found that they have almost opposite directions, see figure \ref{F3Umean}(\textit{b}).   (For $N\to\infty$, the ensemble average of the cross-stream components would tend to be zero.)
Thus these forces merely target particular vortices of the velocity ICs, which indicates that they are an important part of the SSP to target initially, but not a good target for a steady force. 
       
\begin{figure}
       \centering
       \includegraphics[angle=0,width=1\textwidth]{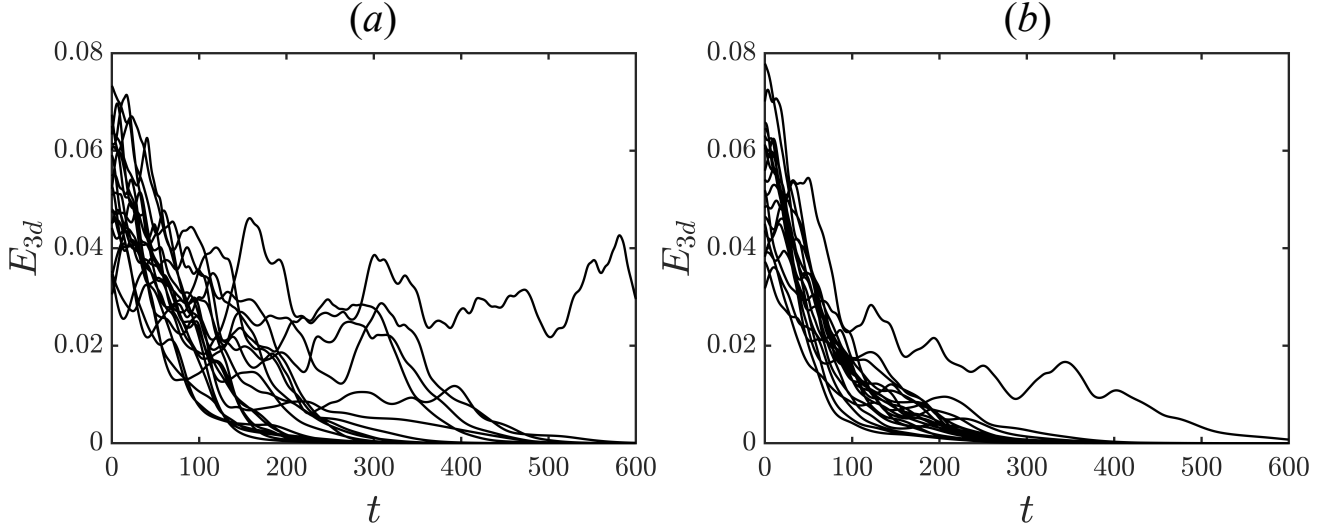}  
       \caption{The body force $\vec F_1$ is applied to another 20 ICs at (\textit{a}) Re=2400, $A_{crit}=5.8e-06$, (\textit{b}) $Re=3000$, $A_{crit}=1.63e-05$ } 
       \label{Robustness-F3-Re2430}
    \end{figure}

\begin{figure}
       \centering
       \includegraphics[angle=0,width=1\textwidth]{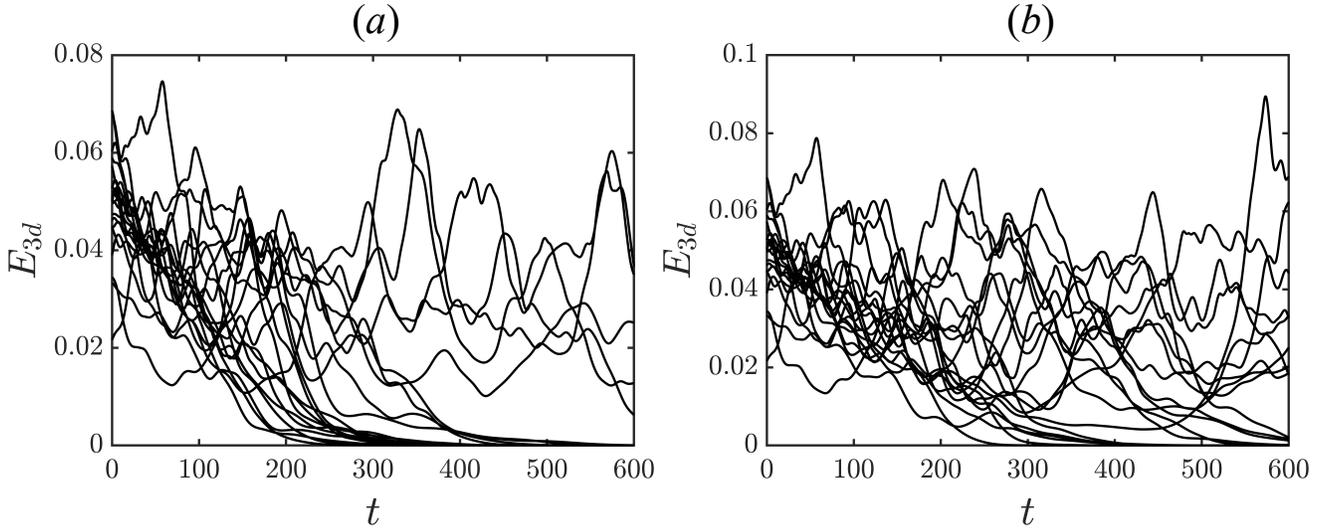}  
       \caption{
       The body force $\vec F_2$ with different force amplitude (\textit{a}) $A_0=7e-6$, (\textit{b}) $A_0=4e-6$ (the critical force amplitude for $\vec F_2$), are applied to another 20 ICs at $Re=2400$}
       \label{F2-check}
    \end{figure}

    To check the robustness of the minimal body force, we applied the optimised body forces $\vec{F}_1$ and $\vec{F}_2$ to $20$ ICs (different from the ICs used in the optimisation).  Time series of $E_{3d}$ for these ICs are shown in figure \ref{Robustness-F3-Re2430}  for $\vec{F}_1$, and figure \ref{F2-check} for $\vec{F}_2$.  At $Re=2400$, although several ICs do not decay to the laminar state by $t=300$,  almost all ICs successfully laminarise within $t=600$, and only one IC failed.   At $Re=3000$, all ICs decay to the laminar state by $t=600$.  These results suggest a good robustness of $\vec F_1$.  For  $\vec{F_2}$, robustness is lost, especially at smaller force amplitude, 
    as some of the budget for the force is misplaced targeting cross-stream components of the ICs used in the optimisation, which differ for these ICs.  Instead, the cross-stream components of the force sometimes strengthen the lift-up process, rather than weakening it.  As a result, many ICs fail to laminarise.   
      
\begin{figure}
          \centering
          \includegraphics[angle=0,width=1\textwidth]{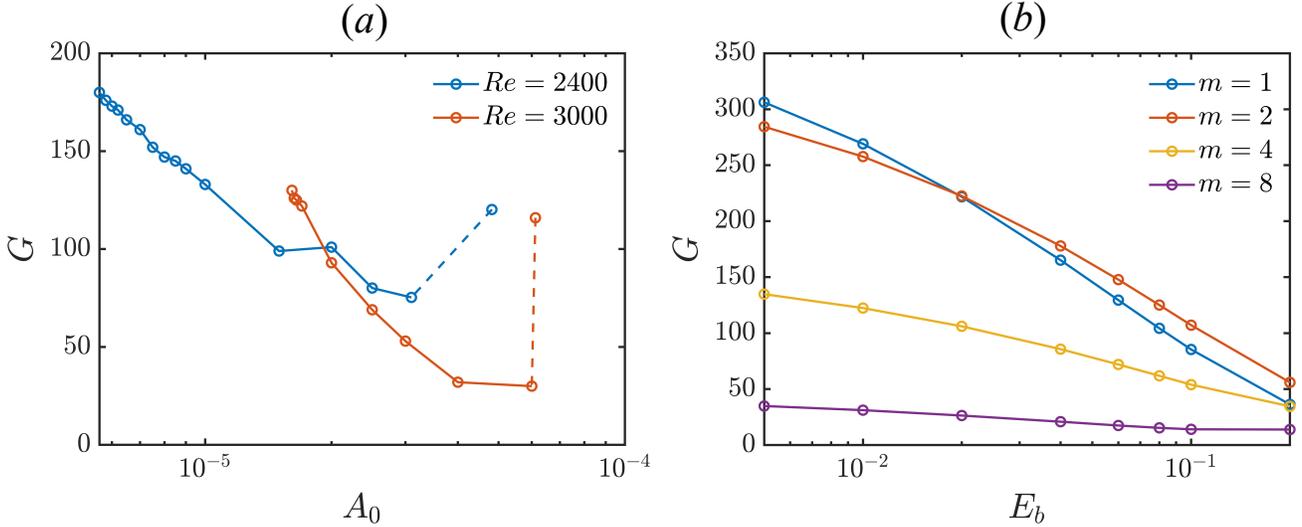}  
          \caption{(\textit{a}) TG  (for $m=1$ perturbations) of the forced laminar profile, the point of largest force amplitude corresponds to the TG of initial guess $\vec F_s$, the dashed line indicates the reduction of TG in the optimisation for the initial guess $F_s$ ($G={<u_{T}^{2} > }/{<u_{0}^{2} >}$). The last point on the left for each curve corresponds to minimal force.  
          (\textit{b}) The energy growth  changes with $E_b$ for different azimuthal wave numbers at $Re=2400$, $G_{max}=403$ when no body force exists.}
          \label{TG-forcedUz}
        \end{figure}

    The comparison of $\vec{F}_1$ and $\vec{F}_2$  together suggests that the streamwise body force is most efficient for laminarisation in the time-independent case, but we would like to understand how the laminarisation works. In K\"uhnen $\textit{et al.,}$ \cite{kuhnen2018destabilizing}, it is thought to be the reduced linear TG that causes the collapse of turbulence.  
    Calculations for the linear TG of our 
    forced laminar velocity profile at each
    amplitude are shown in figure \ref{TG-forcedUz}(\textit{a}).  The first point towards the upper-right is the TG of the initial unoptimised force $\vec F_s$.  
    For the first optimisation, a substantial reduction of TG is observed.  Then the TG increases gradually as the force amplitude is decreased.  When the TG exceeds a certain level, turbulence cannot be eliminated.  
        
\section{Linear optimisation} 
\label{sec:headings} 

    After an expensive nonlinear optimisation, the observations at the end of the last section suggest that minimisation of linear TG may be a much simpler and numerically cheaper means to determine an optimal forced laminar profile and corresponding force. In this section, we examine whether it's really that straightforward.

    \subsection{Formulation}  
    In order to optimise the forced laminar profile, we  decompose the total velocity and pressure fields as $\vec{u}_{tot}=U \hat{\vec z}+\vec {u}_b +\vec{u}$ and $p_{tot}=P+p_b+p$, with
    $\vec{u}_b=(0,0,u_{b,z}(r))$,    
    where  $U = 1-r^2$  is the parabolic laminar profile,
    $U+{u}_{b,z}$ is the forced laminar profile,  
    and  $\vec{u}=(u_r,u_\theta,u_z)$ and $p$ are the deviations from the forced laminar fields. 
Here we consider optimisation of 
only $u_{b,z}(r)$ for simplicity.  
    Via the decomposition, the amplitude of 
    $u_{b,z}$ can be constrained in an optimisation, and may be considered as a proxy for amplitude of the body force used to shape the forced laminar profile.  The Navier--Stokes and continuity equations for the perturbations are
       \begin{equation}
       	\frac{\partial \vec{u}}{\partial t} + U\frac{\partial \vec u}{\partial z}+U^{'}u_r\hat{\vec z} -\vec u \times \vec{\nabla} \times \vec u -\vec u_b \times \vec \nabla \times \vec u-\vec u\times \vec \nabla \times \vec u_b  =\frac{1}{Re} \vec{\nabla}\vec{u}^{2}-\vec{\nabla} p
       \end{equation}
       \begin{equation}
       	\vec{\nabla} \boldsymbol{\cdot} \vec{u}=0.
       \end{equation}  
      Here, we derived the full nonlinear equation for completeness. However, in numerical computations, $\vec{u}_0\equiv \vec{u}(t=0)$ is set
      to be small, $E_0=1e-15$, so that calculations are effectively linear.
      The linear equation 
      separates into a set of equations for each azimuthal wavenumber $m$. The initial perturbation that grows the most, i.e.\ the linear optimal perturbation (LOP), could vary with changes in the laminar profile, but the change in a LOP for a given $m$ is expected to be small, provided that inflectional instability is not triggered. In a linearly stable pipe flow system, the LOP of the first azimuthal wave number is one pair of streamwise vortices filling the pipe, while the LOP of the second azimuthal wave number is two pairs of streamwise vortices with great rotation symmetry. 
      To be numerically practical, we fix $\vec{u}_0$ to be the LOP
      for a given $m$ for the parabolic laminar profile.
      We then consider the optimal $u_{b,z}(r)$ that leads to the  minimum TG starting from $\vec{u}_0$. The energy of the perturbation $\vec{u}_0$ at time $t=T$ is minimised with the constraints that it satisfies the three-dimensional Navier-Stokes equations, constant mass flux, a given amplitude of $\left< {\vec u_b} ^{2}\right> =E_b$ and zero flux of $\vec {u}_b$. The target time T is chosen to be just large enough for the occurrence of maximum energy growth.
       The Lagrangian is 
       \begin{eqnarray}
       		\mathcal L=&&\frac{1}{2N} \sum_{n=1}^{N}\left <\vec{u}_n(\vec{x},T_n)^2 \right > +\lambda(\frac{1}{2}\left<{\vec u_b}^{2}\right>-E_{b})+\gamma \left<{\vec u_b \boldsymbol{\cdot} \hat{\vec z}}\right>+\sum_{n=1}^{N}\int_{0}^{T}\left \langle \vec{v}_n\boldsymbol{\cdot} \mathrm{NS}(\vec{u}_n)\right \rangle\mathrm{d}t \nonumber\\
       		&&+\sum_{n=1}^{N}\int_{0}^{T}\left \langle \Pi (\vec{\nabla}\boldsymbol{\cdot} \vec{u}_n) \right \rangle\mathrm{d}t+\sum_{n=1}^{N}\int_{0}^{T} \Gamma (t)\left \langle (\vec{u}_n\boldsymbol{\cdot} \vec{\hat{z}}) \right \rangle\mathrm{d}t.
        \label{LG-TG}
       \end{eqnarray}
    The sum here indicates that we can perform the optimisation for a set of $m$ and their corresponding $\vec{u}_0$,
       or just for a single $m$ and its $\vec{u}_0$ if $N=1$.
       Taking variations of $\mathcal L$ and setting them to be zero, we obtain Euler-Lagrange equations for each $\vec{u}_n$.
       The adjoint Navier--stokes and continuity equations are
       \begin{eqnarray}
       		\frac{\partial \mathcal L} {\partial \vec{u}_n}=&&\frac{\partial \vec{v}_n}{\partial t}+U\frac{\partial \vec{v}_n}{\partial z}-v_{z}{U}'\vec{\hat{r}} +\vec{\nabla}\times(\vec{v}_n\times \vec{u}_n)- \vec{v}_n\times\vec{\nabla} \times \vec{u}_n\nonumber\\
         &&+\vec{\nabla}\times(\vec{v}_n\times \vec {u}_b)- \vec{v}_n\times\vec{\nabla} \times \vec{u}_b +\nabla \Pi + \frac{1}{Re}\nabla ^{2}\vec{v}_n-\Gamma\vec{\hat{z}}=0,
       \end{eqnarray}
       \begin{equation}
       	\frac{\delta \mathcal L}{\delta p_n}=\vec \nabla \boldsymbol{\cdot} \vec{v}_n =0.
       \end{equation}
       The compatibility condition (terminal condition for backward integration) is given by
       \begin{equation}
       \label{eq:vt}
       	\frac{\delta \mathcal L}{\delta \vec{u}_n(\vec{x},T)}=\vec{u}_n(\vec{x},T_n)+\vec{v}_n(\vec{x},T_n)=0
       \end{equation} 
       and optimality condition
       \begin{equation}
       	\frac{\delta \mathcal L}{\delta \vec u_b}=\lambda \vec {u_b} +\gamma +\mathcal{F}({\vec \sigma})=0,
       \end{equation}
       where \begin{equation}
       	{\vec \sigma}(r,\theta, z)=\frac{1}{N} \sum_{n=1}^{N}\int_{0}^{T}  (\vec{v}_n\times\vec{\nabla} \times \vec {u}_n -\vec{\nabla}\times(\vec{v}_n\times \vec {u}_n ))\mathrm{d}t
       \end{equation}
       is a vector function of space. Unwanted spatial dependence of azimuthal and streamwise direction is removed by the Fourier filter $\mathcal{F}$ which ensures that the update to $\vec{u}_b$ is 
       dependent on $r$ only.
       $\gamma$ is a Lagrangian multiplier 
       that ensures zero flux of $\vec {u}_b$, determined by
       \begin{equation}
       \left <\left |\frac{\delta \mathcal L}{\delta \vec u_b} \boldsymbol{\cdot} \hat{\vec z} \right | _{k,m=0} \right >= \left <\left |\lambda\vec u_{b,z}+\gamma+\mathcal{F}\sigma_z \right |_{k,m=0} \right >=0.
       \end{equation}
       The update for $\vec{u}_b$ is 
       \begin{equation}
       	\vec {u}_b^{(j+1)}=\vec{u}_b^{(j)}-\epsilon\frac{\delta \mathcal L}{\delta\vec {u}_b^{(j)}}
      \qquad \mbox{i.e.} \qquad 
       	u_{b,z}^{(j+1)}=u_{b,z}^{(j)}-\epsilon\left.\frac{\delta \mathcal L}{\delta\vec {u}_b^{(j)}}\boldsymbol{\cdot} \hat{\vec{z}}
        \right|_{k,m=0}   
       \end{equation}
%
       After finding the optimal forced laminar flow, the body force that gives this forced laminar flow can be calculated by evaluation of $z$-component of the mean equation
       \begin{equation}
       	\frac{1}{Re}\nabla^2 \vec{u}_b + F(r)\hat{\vec{z}}=0\,. 
       \end{equation}
       
       The initial input $\vec u_{b}^{(0)}$ is arbitrary, provided that zero flux is met.  
       As before, $Re=2400$ and $Re=3000$ are considered with $L=10$.  
       Given the linearisation and symmetry, the 
       resolution may be reduced to
       $S=64, M=10, K=1$, and the time step is $\Delta t=0.01$. The LOPs for azimuthal wave numbers $m=1,2,4,8$ are adopted as ICs to find the optimal forced laminar flow which can suppress TG at these wave numbers, respectively. 
  
  \subsection{Results for linear optimisation} 
     \subsubsection{$Re=2400$}  
\begin{figure}
      	\centering
      	\includegraphics[angle=0,width=1\textwidth]{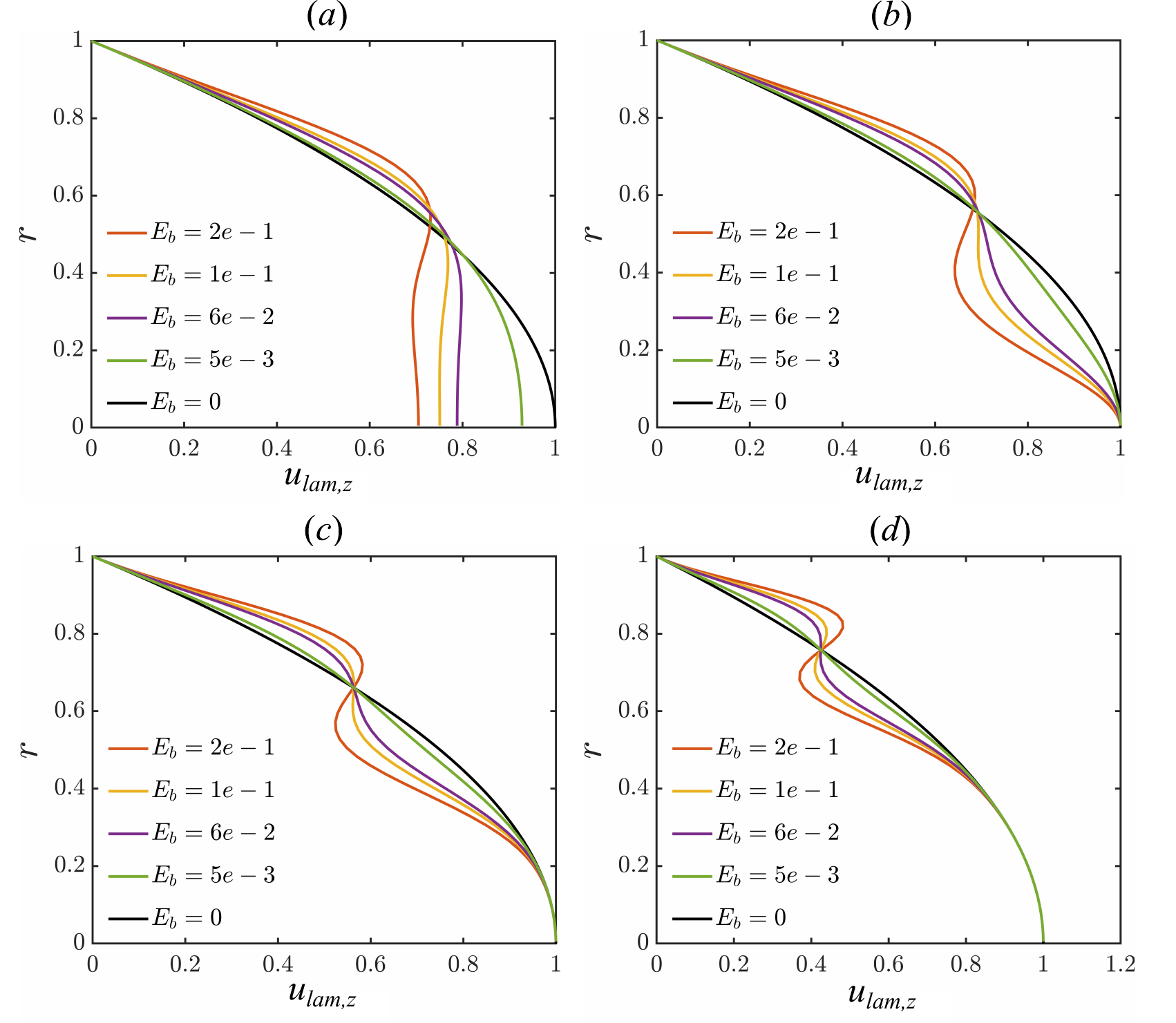}  
      	\caption{The optimal forced laminar profile for several $E_b$ at $Re=2400$: (\textit{a})$m=1$, (\textit{b})$m=2$ (\textit{c})$m=4$, (\textit{d})$m=8$.}
      	\label{Prof-Eb-m}
      \end{figure}

       The minimised energy growth of each azimuthal wave number versus several $E_b$ is presented in figure \ref{TG-forcedUz}(\textit{b}). Compared with the unforced parabolic laminar profile, the TG of the optimal profile decreases substantially, especially for lower azimuthal wavenumber $m$.  Figure \ref{Prof-Eb-m} shows the optimal laminar velocity profile for several $E_b$  after optimising separately ($N=1$) for each of four azimuthal wave numbers. The optimisation reduces the radial velocity gradient in a specific radial interval for each $m$, but the velocity gradient near the wall increases for all cases. The radial interval of the reduced gradient is closer to the wall for higher $m$, corresponding to the location of the roll in the (LOP) IC. It confirms that the radial velocity gradient plays a dominant role for TG \citep{brandt2014lift}.  The flattened forced laminar profile  at the centre is only observed for the case $m=1$. This form of flattened profile is quite similar to the profile found by Hof $\textit{et al.,}$ \cite{hof2010eliminating} and K\"uhnen $\textit{et al.}$ \cite{kuhnen2018destabilizing}, which can laminarise turbulence.  At larger $E_b=2e-1$, there is a clear negative velocity gradient, which brings inflectional points to the velocity profile and causes linear instability. The forced laminar profile for other $m$ all exhibit inflections, indicating that the corresponding body force may not be able to eliminate turbulence. 
      
\begin{figure}
       	\centering
       	\includegraphics[angle=0,width=1\textwidth]{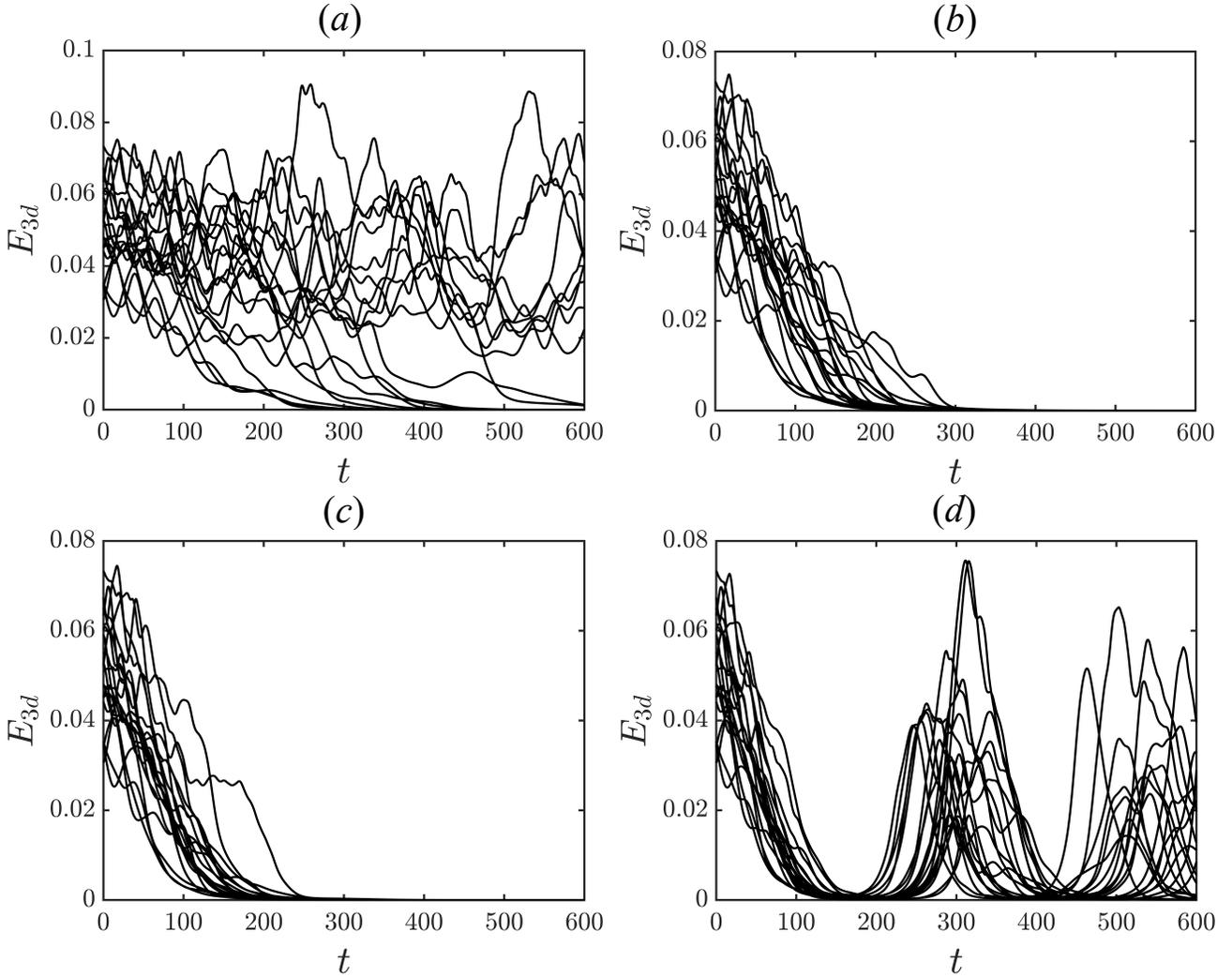}  
       	\caption{$E_{3d}$  of the turbulence with corresponding body force of different optimal forced laminar flow $\vec{u}_b$ : (\textit{a}) $E_b=5e-3$, (\textit{b}) $E_b=6e-2$, (\textit{c}) $E_b=1e-1$, (\textit{d}) $E_b=2e-1$, the simulations are run at $Re=2400$, and the optimal forced laminar flow is optimised at $m=1, Re=2400$.}
       	\label{Robustness-Ub}
       \end{figure}

       To determine the effect of reduced linear TG on turbulence, we introduced the corresponding body force into 20 simulations with random initial states at $Re=2400$, picked from a turbulent simulation with a  time interval $\Delta t=50$.  Given the expected inflectional instability in the optimised forced laminar flows for $m>1$, we examine the $m=1$ case only. Results are shown in figure \ref{Robustness-Ub}. When $E_b$ is small, TG is reduced only a little, and turbulence survives in many cases (figure \ref{Robustness-Ub}(a)).  For a larger $E_b$, all turbulent ICs decay to the laminar state (figure \ref{Robustness-Ub}(b)). When $E_b$ is increased further, turbulence decays more quickly (figure \ref{Robustness-Ub}(c)).  But when there is a negative velocity gradient,  temporal intermittency is observed. This may be caused by the appearance and disappearance of linear instability, as the linear instability triggers turbulence that flattens the velocity profile, which suppresses the linear instability, but cyclically also leads to the collapse of the  turbulence \citep{chu2024minimal}. 
       These results indicate the significant role of TG in sustaining shear turbulence in linearly stable flows \citep{kim2000linear}.

    \subsubsection{$Re=3000$} 
\begin{figure}
       \centering
       \includegraphics[angle=0,width=1\textwidth]{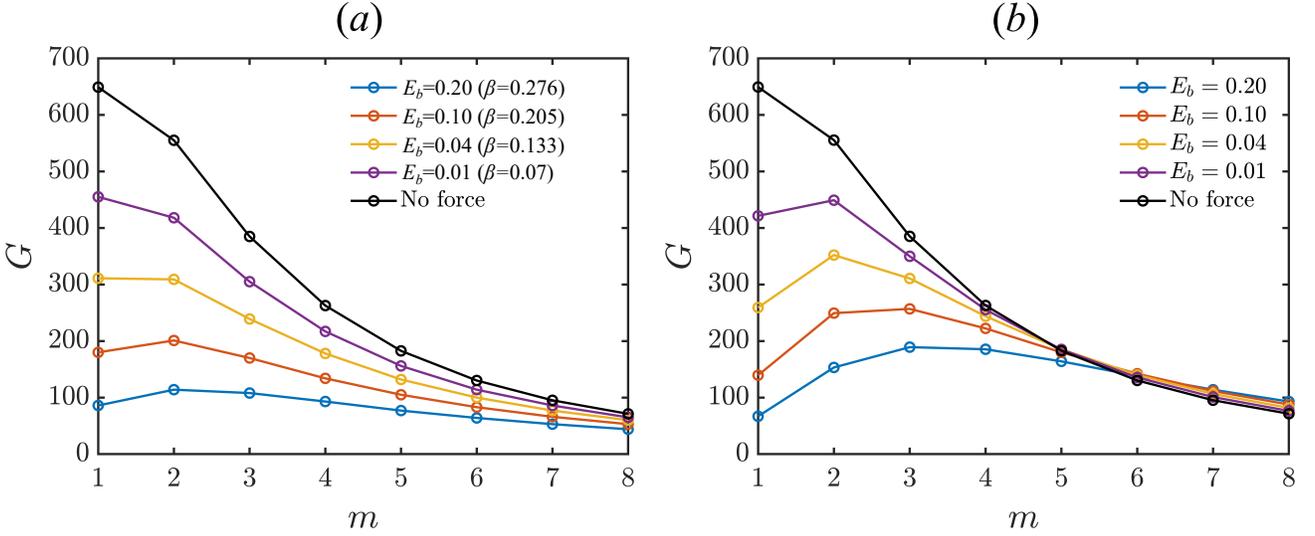} 
       \caption{TG of different $m$ for (\textit{a}) the forced laminar profile constructed by Song $\textit{et al.}$ \cite{song2014direct} for different values of $\beta$, (\textit{b}) the forced laminar profile obtained by optimisation for $m=1$ only at
       $Re=3000$.
       }
       	\label{SongTG}
       \end{figure}

    $Re=2400$ is near the critical Reynolds number for sustained turbulence.  To determine whether this approach works at a higher Reynolds number, we calculated the optimal forced laminar flow at $Re=3000$ and the corresponding body force,  still using $m=1$ only in the linear optimisation. Simulations with $20$ turbulent ICs were then performed at $Re=3000$. This time, the laminarisation is not so promising as figure \ref{Robustness-Ub}, and the turbulence of several ICs failed to decay, even when the forced laminar profile was quite flattened.  We compare the TG for our profiles with the forced laminar profile designed by Song $\textit{et al.}$ \cite{song2014direct}, which was shown to laminarise turbulence up to $Re=10^5$. The TG of the first eight azimuthal wave numbers for two forced laminar profiles is shown in figure \ref{SongTG}. The key difference that can be observed is that the forced laminar profile of Song $\textit{et al.}$ \cite{song2014direct} reduces the TG of all first eight modes, rather than (optimally) reducing just $m=1$. It has a great suppression of TG for lower wave numbers $m<4$,  while our optimal forced laminar flow focuses on reducing the TG of $m=1$, the TG of $m=2$ is still considerable. The reduction of TG of $m=1$ is equivalent to a decrease in the radial velocity gradient of the specific radial interval (near centre), see figure \ref{Prof-Eb-m}($a$). The focus of reducing the TG of $m=1$ will increase the radial velocity gradient more near the wall for fixed flux; thus the TG of $m=2$ and the higher wave number are higher than that of the forced laminar profile designed by Song $\textit{et al.}$ \cite{song2014direct}. 
    This observation suggests that optimisation may need to  reduce TG at multiple $m$ simultaneously.

    As the required reduction in TG is not expected to be the same for each $m$, we introduce a new target function 
    \begin{equation}
        	\mathcal T =\sum_{n=1}^{N}\omega_{n}\langle \vec{u}_n(\vec{x},T_n)^2 \rangle\,, 
         \label{eq:target-function}
        \end{equation}
    where $\omega_n$ is a scalar weight and 
    $T_n$ is a target time for each IC.
    The values of $\omega_n$ are adjusted
    to balance the growth for each wavenumber.
    The Lagrangian equation is similar to equation \eqref{LG-TG}, but the initial condition of the adjoint equation 
    is different:
        \begin{equation}
        	\vec{v_n}(T_n)=-\omega_{n} \vec{u_n}(T_n).
        \end{equation} 
    \begin{figure}
        	\centering
        	\includegraphics[angle=0,width=1\textwidth]{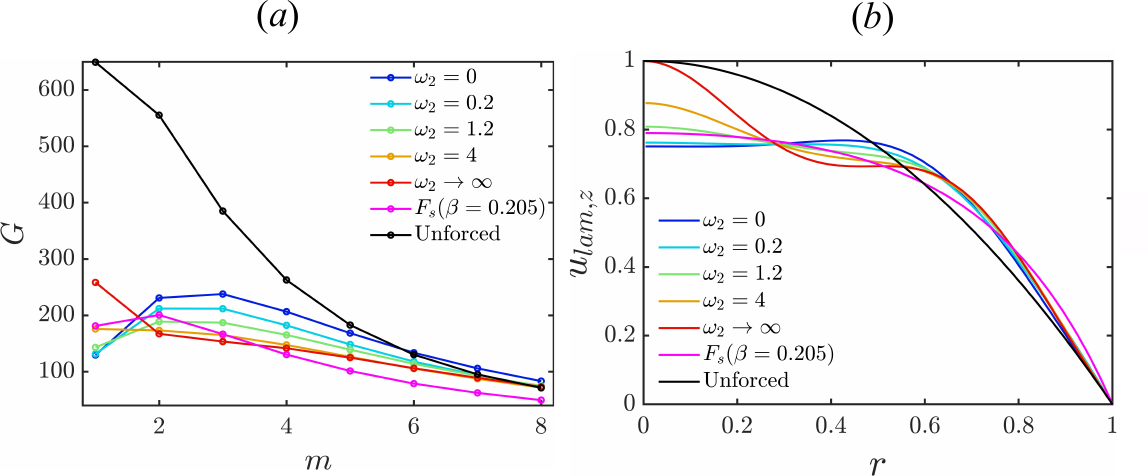}  
        	\caption{(\textit{a}) 
         The TG of each wave number for the profile obtained when minimising TG of the first
         two wavenumbers $m=1$ and $m=2$,
         for several weights $\omega_2$ and fixed $\omega_1=1$ and $E_b=0.1$.  Also plotted are TGs for the profile of Song $\textit{et al.}$ \cite{song2014direct}, $\vec{F}_s$; (\textit{b})  comparison of corresponding optimised forced laminar profiles. 
         }
        	\label{TG-Eb-m12}
    \end{figure}
        
    We first minimise the TG of the first two azimuthal wave numbers, as the TG of $m=2$ is still considerable. We gradually increase the weight function of the second mode from $\omega_2 = 0$ (with fixed $\omega_1=1, E_b=0.1$) for each optimisation to reduce the TG of the second azimuthal wave number more, and test whether it can laminarise the turbulence at $Re=3000$.  The TG of the forced laminar profile with increasing $\omega_2$, and the corresponding forced laminar profiles, are shown in figure \ref{TG-Eb-m12}.  
    Compared with the TG when only
    $m=1$ was optimised,
    there is the reduction of the TG 
    for $m=2$ as expected, but also considerable reduction for higher wave numbers too. 
    Increasing $\omega_2$ from $0$ to `infinity' while keeping $\omega_1=1$ corresponds to switching the balance from optimising only $m=1$ to optimising only $m=2$.
  When $\omega_2$ is large (from approximately 1.2), a stronger inflection point is observed in the profile.  The profile of Song $\textit{et al.}$ \cite{song2014direct} for the same $E_b$ ($\beta=0.205$) is capable to reduce the TG at higher wave numbers more than the other cases, while keeping the growth relatively small also at $m=1$ and $m=2$.  

\begin{figure}
        \centering
        \includegraphics[angle=0,width=1\textwidth]{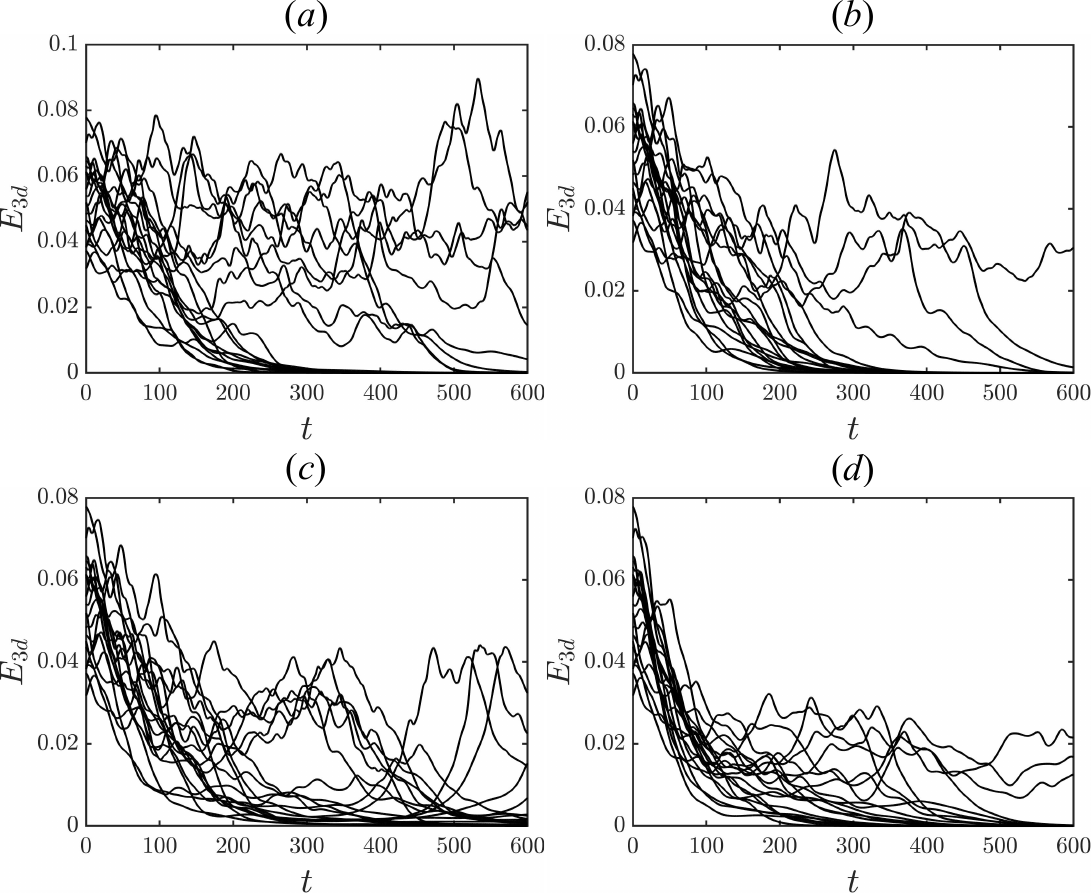}  
        \caption{The laminarisation tests for $20$ turbulent ICs at $Re=3000$ when different forced profile is adopted: 
        (\textit{a}) $\omega_2=0$; (\textit{b}) $\omega_2=0.2$; (\textit{c}) $\omega_2=1.2$; (\textit{d}) ${F}_s(\beta=0.205)$.}
        \label{Robustness-Re3000-omega}
        \end{figure}

  The laminarising effect of these body forces on turbulence are shown in figure \ref{Robustness-Re3000-omega}.  The body force obtained when optimising the TG of only $m=1$ can still laminarise some ICs, but fails to laminarise all ICs (figure \ref{Robustness-Re3000-omega}(\textit{a})). When we include minimisation of TG of $m=2$, the corresponding force of the optimised profile can laminarise almost all ICs again (figure \ref{Robustness-Re3000-omega}(\textit{b})) --- compared to the case at $Re=2400$, laminarisation at the higher Reynolds number requires reduction of TG of more than just the first wavenumber $m=1$.  Too much focus on $m=2$ (too larger $\omega_2$) brings strong inflectional instability, thus full decay of turbulence is not possible, although $E_{3d}$ can decay to quite low values.   The body force of \citep{song2014direct} can laminarise most turbulence, but not as well as the optimised profile for $\omega_2=0.2$.  It should be noted that the force amplitude of $F_s(\beta=0.205)$ is 3.31e-05, while that of our force with $\omega_2=0.2$ is 9.98e-06, although they have the same $E_b=0.01$. 
  The large difference is due to the fact that $F_s(\beta=0.205)$ consumes a lot of energy to reduce the TG more at all higher wave numbers, which is not necessary for laminarisation
  of turbulence at $Re=3000$.  

\begin{figure}
        \centering
        \includegraphics[angle=0,width=1\textwidth]{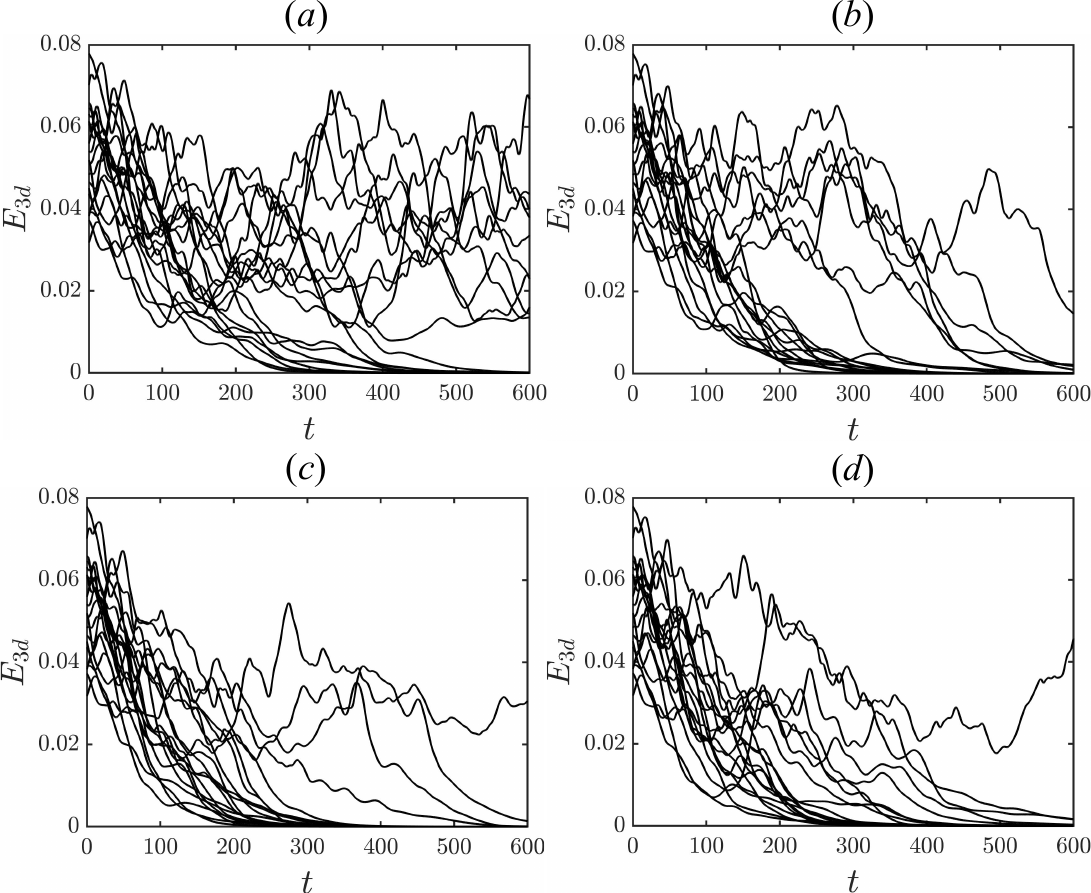}  
        \caption{The laminarisation tests for $20$ turbulent ICs at $Re=3000$ 
        (\textit{a}) $E_b=6e-2$; (\textit{b}) $E_b=8e-2$; (\textit{c}) $E_b=1e-1$; (\textit{d}) $E_b=1.2e-1$ 
        }
        \label{Robustness-Re3000}
        \end{figure}

   The optimal profiles at different $E_b$ using $\omega_1=1\,, \omega_2=0.2$ are also calculated. The corresponding body forces are introduced to another 20 simulations with turbulent ICs to check whether they can eliminate turbulence, as shown in figure \ref{Robustness-Re3000}. 
   For a small $E_b$, it is observed that most  turbulence survives, but with a larger $E_b$, most cases decay again.  Similar laminarisation can be also found at larger $E_b$. These results support the conjecture that the laminarisation at a higher Reynolds number requires reductions of TG for a higher azimuthal wave number.   

    Overall, it has been shown that turbulence at $Re=2400$ and $Re=3000$ can be eliminated by reducing the TG of forced laminar flow. However, it is found to be very hard to eliminate turbulence at higher Reynolds numbers. For example at $Re=4000$, we need to minimise the TG in the first ten azimuthal wave numbers. 
    In addition, it should be stated that it is not easy to determine how much the TG at each $m$ should be reduced (i.e. how to choose the weights in \eqref{eq:target-function}). 
    Overall, it is not so straightforward to use linear optimisation alone to eliminate turbulence at a high Reynolds number.    
        
\section{Discussion}
     \label{sec:headings} 
   \subsection{Mechanism of laminarisation}  
     An important part to discuss is the mechanism of turbulence laminarisation caused by flattening the forced laminar profile. 
     It is an exciting but not surprising result that the turbulence can be eliminated by reducing TG, e.g.\  Kim $\textit{et al.,}$ \cite{kim2000linear} have investigated the linear process in wall-bounded turbulent shear flow, and found that the linear coupling term, which enhances TG of linearized Navier-Stokes system, plays a significant role in maintaining turbulence. Near-wall turbulence is shown to decay without the linear coupling term. 
    Marensi $\textit{et al.,}$ \cite{marensi2019stabilisation} researched the minimal seed for transition in the presence of a flattened forced laminar profile and found that the flattened forced laminar profile can reduce the energy growth of the minimal seed, namely enhancing the nonlinear stability of the forced laminar profile. 
     From the perspective of the 
     production and dissipation of turbulent energy, Budanur $\textit{et al.,}$ \cite{budanur2020upper} observed that the transition to turbulence relies on energy amplification away from the wall. Thus turbulence decays when the forced laminar profile is flattened in the center. 
     
\begin{figure}
      	\centering
      	\includegraphics[angle=0,width=1\textwidth]{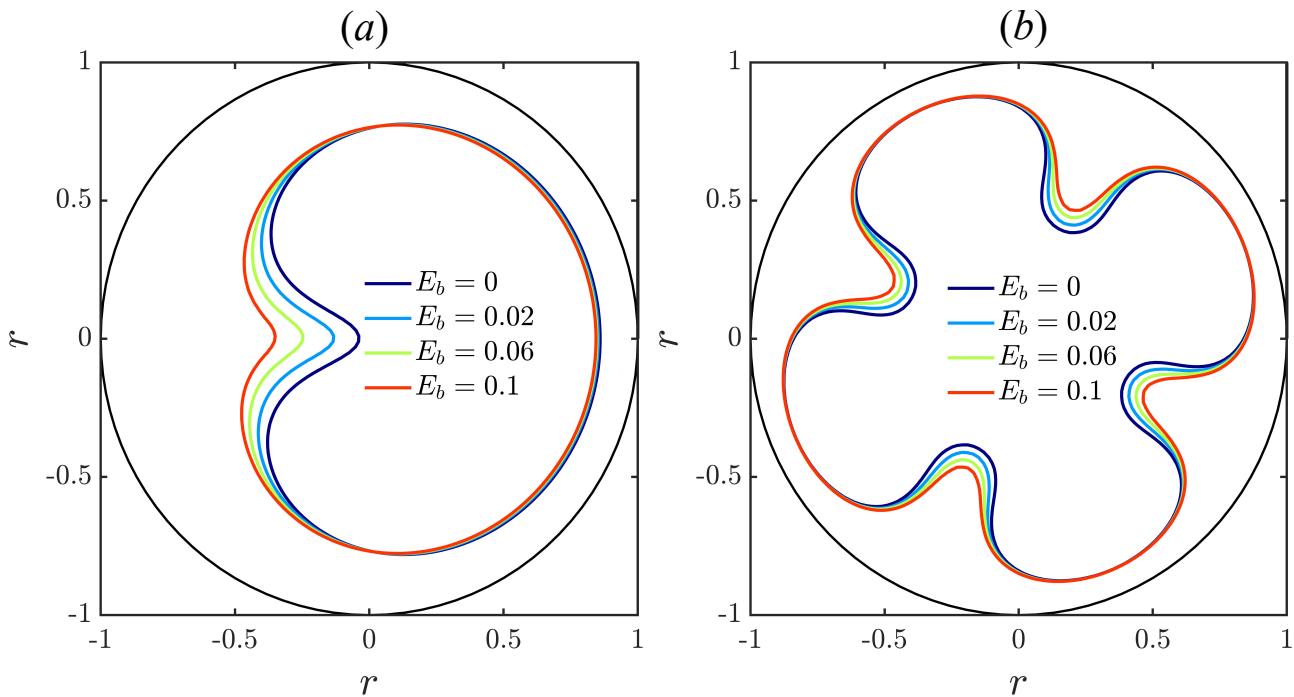}  
      	\caption{The maximum streaks formed by LOP of (\textit{a}) $m=1, E_0=3e-3$, (\textit{b}) $m=4, E_0=3e-3$ with the flattened forced laminar profile. The contour is marked by $u_z=0.55, 0.45$, respectively. The flattened profile is obtained by linear optimisation at Re=2400}
      	\label{streaks-LOP}
      \end{figure}

    K\"uhnen $\textit{et al.,}$ \cite{kuhnen2018destabilizing} argued that the reason for laminarisation is the reduction of optimal TG and suggested a critical TG for laminarisation. However, according to our results,  we add that the reduction needs to target a particular range of azimuthal wave numbers for turbulence to be eliminated. 
     In figure \ref{streaks-LOP}, we plot the contour of streaks formed by the LOP for several azimuthal wave numbers with the flattened forced laminar profile.  For larger $E_b$,  meaning more reduced TG, the lifted low-speed streaks are closer to the wall in each azimuthal wave number.  The streaks closer to the wall are usually more stable according to Schoppa $\textit{et al.,}$ \cite{schoppa2002coherent}, and cannot produce streak instability \citep{hamilton1995regeneration,jimenez1999autonomous} or streak TG (also called secondary TG) \citep{schoppa2002coherent,cossu2007optimal,lozano2021cause} to sustain turbulence. The TG is produced by the non-orthogonality of the eigenvectors of the laminar solution, while the streak TG (secondary TG) refers to rapid energy growth caused by the  non-orthogonality of the eigenvectors of the streaks. The reduction of TG  at each azimuthal wave number leads to suppressing the formation of streaks of these wave numbers. Turbulence at a higher Reynolds number can be sustained by streaks of larger azimuthal wave numbers, which can be inferred from the decrease of the minimal domain for sustained turbulence at higher Reynolds numbers \citep{jimenez1991minimal}. Thus we need a reduction of TG at more azimuthal wave numbers to suppress the formation of streaks of smaller azimuthal wavelengths to laminarise the turbulence at higher Reynolds numbers.  
     
\begin{figure}
     	\centering
     	\includegraphics[angle=0,width=1 \textwidth]{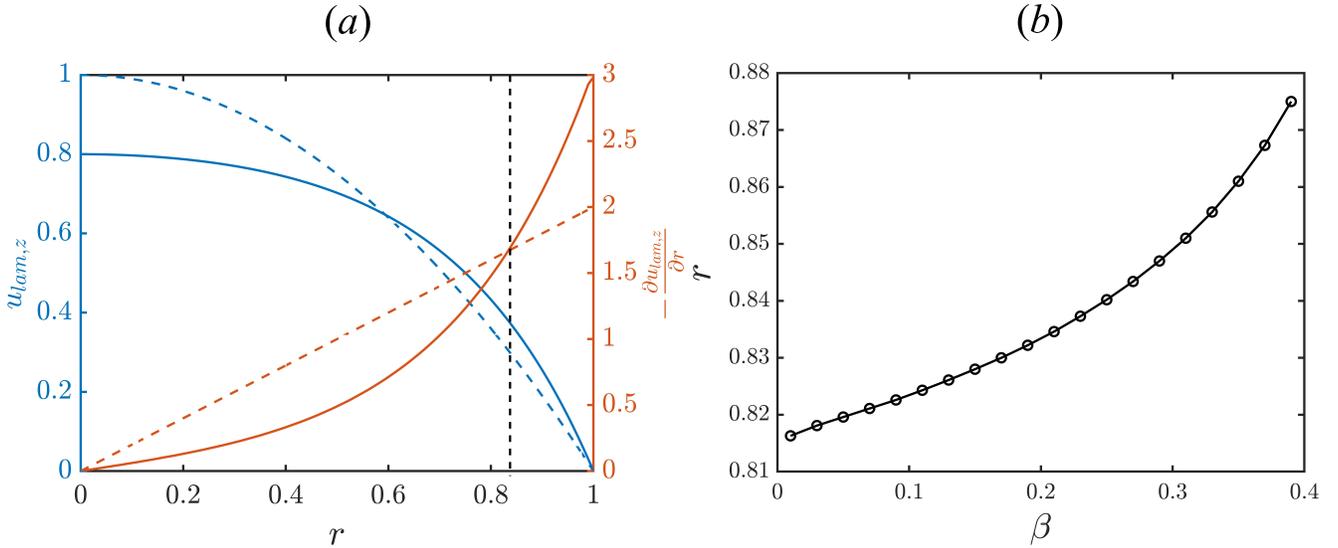} 
     	\caption{$(a)$ The radial velocity gradient of parabolic velocity 
             profile and flattened velocity profile. The dotted line means parabolic velocity profile, and solid line represents the flattened forced laminar profile \citep{song2014direct}$(\beta=0.2)$. The dash black line represent the position where the gradient of the parabolic profile equals to that of the flattened profile.  $(b)$ The position where the gradient of the parabolic profile equals to that of the flattened profile of Song $\textit{et al.}$ \cite{song2014direct} at different $\beta$. }
     	\label{gradient}
     \end{figure}

     We can also probe the mechanism focussing on the radial velocity gradient, which is of great significance for TG. 
     The radial velocity gradient of parabolic velocity and flattened velocity profile $(\beta=0.2)$ designed by Song $\textit{et al.}$ \cite{song2014direct} are plotted in figure \ref{gradient}$(a)$. When the velocity profile is flattened, it can be found that the velocity gradient in the center decreases, while the velocity gradient near the wall increases.  As a result, the regeneration of streaks in different sizes  will be affected.  Streaks smaller than the vertical dotted black line are enhanced while streaks taller than the dotted black line are suppressed. 
     When the forced laminar profile is flattened enough, no unstable streaks are generated, and then the turbulence cannot be sustained by streak instability or streak TG. The crossing point where the radial gradient of the parabolic profile equals that of the flattened profile is plotted in figure\ref{gradient}$(b)$ for different $\beta$. The position of the crossing
     is closer to the wall with a larger $\beta$, consequently, the profile of larger $\beta$ can laminarise the turbulence at higher Reynolds numbers \citep{kuhnen2018destabilizing} due to the suppression of smaller (i.e. closer to the wall) streaks. It should be noted that although in our previous calculations we applied the body force to a turbulent profile, we believe the fundamental mechanism will be the same for the laminar solution. The body force changes the dynamics around the laminar solution, i.e. enhancing the nonlinear stability of the laminar state \citep{marensi2019stabilisation}, leading to a collapse of turbulence. Therefore, here we use the laminar profile to analyse the TG rather than the turbulent mean profile.
     
\begin{figure}
      	\centering
      	\includegraphics[angle=0,width=1\textwidth]{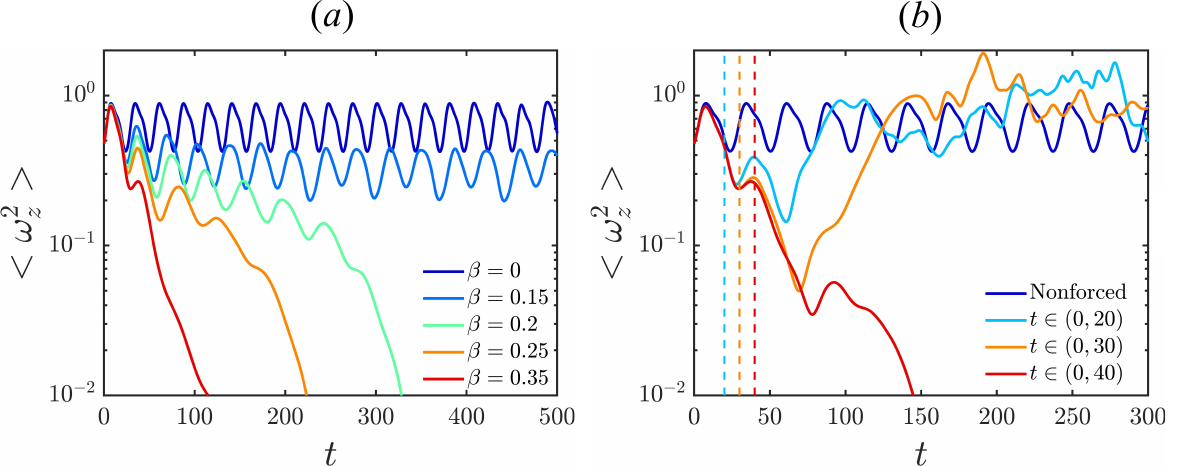}  
      	\caption{The time evolution of the amplitude of streamwise vorticity $\langle\omega_z^2 \rangle$  when the relative periodic orbit is forced by Song's force \citep{song2014direct}. $(a)$ The force of different $\beta$ is imposed all time; $(b)$ the  force of $\beta=0.35$ is imposed for different periods of time, and three dash lines mean $t=20, 30, 40$, respectively. The relative periodic orbit is found at $Re=2500, L=3.7, m_p=4$ by Budanur $\textit{et al.}$ \cite{budanur2017relative}. 
       }
      	\label{Po-energy-beta}
      \end{figure}

     A simple numerical experiment was carried out to verify our conclusion about the mechanism of laminarisation. It is well known that the shear turbulence is driven by the SSP \citep{hamilton1995regeneration,waleffe1997self}, which consists of streak formation (linear lift-up process), streak breakdown (instability) and streamwise vortices regeneration (nonlinear interactions). If we can reproduce this typical SSP, and impose the body force on these three stages, respectively, to see whether the SSP can be disrupted, then perhaps we can confirm which stage is suppressed by the body force. 
A reproducible way to analyse the SSP is via relative periodic orbits \citep{kawahara2001periodic}, which have been found in pipe flow \citep{duguet2008relative,willis2013revealing,budanur2017relative}. Here we use the relative periodic orbit found at $Re=2500, L=3.7, m_p=4$ in Budanur $\textit{et al.}$ \cite{budanur2017relative}, where $m_p=4$ indicates that 4-fold rotational symmetry was imposed. The period of the relative periodic orbit is $T_{rpo}=25.65$. Figure \ref{Po-energy-beta}$(a)$ shows the changes in the amplitude of streamwise vorticity when the relative periodic orbit  is forced by Song's force \citep{song2014direct} at different $\beta$. A larger $\beta$ means a more flattened forced velocity profile. When $\beta=0$, there is no force, and the  flow evolves periodically.  The small $\beta=0.15$ fails to laminarise the relative periodic orbit, while $\beta=0.2$ can make it, although requiring some time. A larger $\beta$ can achieve a quicker laminarisation. Then, we tested how much time is needed for laminarisation with the body force of $\beta=0.35$, see figure \ref{Po-energy-beta}$(b)$. 
It can be seen that we need to impose the body force up to at least $T=40$ for laminarisation. When the body force works for less than $40$ time units, the drop of streamwise vorticity can also be observed, but it rises  again after the force is removed. 
        
\begin{figure}
      	\centering
      	\includegraphics[angle=0,width=1\textwidth]{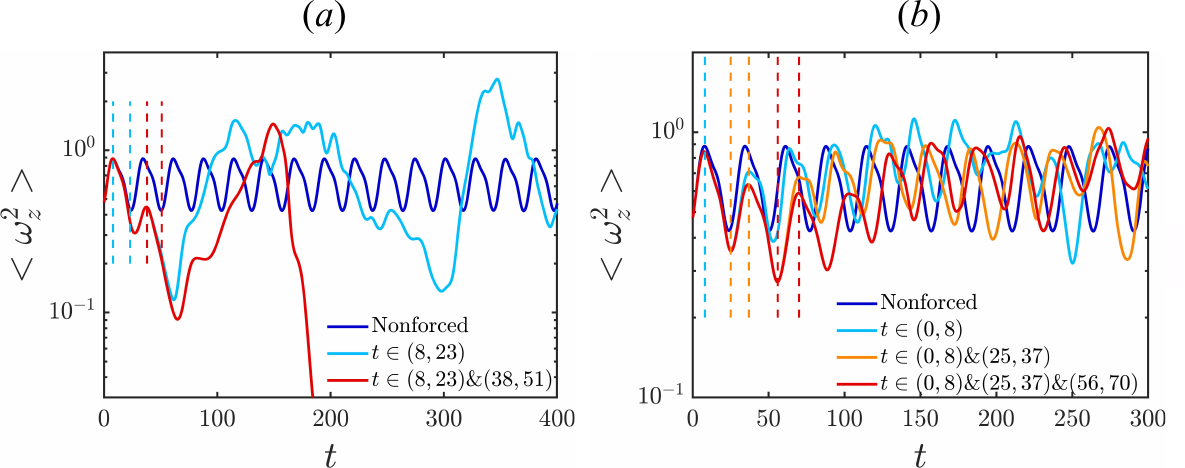}  
      	\caption{The time evolution of the amplitude of streamwise vorticity $\langle\omega_z^2 \rangle$ when the relative periodic orbit is forced by Song's force \citep{song2014direct} $(\beta=0.35)$ \citep{song2014direct} in different stages. $(a)$ Streamwise vorticity becomes weaker, indicating the formation of streaks. The dash lines correspond to $t = 8,23,38,51$ respectively. $(b)$ Streamwise vorticity becomes stronger, meaning the regeneration of streamwise vortices. The dash lines represent $t=8,25,37,56,70$ respectively.}
      	\label{Po-energy-Tf}
      \end{figure}

     As the force is constant, we still cannot ensure which part is disrupted. 
     Hence, we impose the body force of $\beta=0.35$ on the three stages of SSP, respectively. The drop of streamwise vorticity amplitude corresponds to the formation of streaks.  The rise of streamwise vorticity amplitude represents the  regeneration of streamwise vortices.  The stage of streak breakdown occurs quickly and is hard to detect \citep{schoppa2002coherent}, thus the effect of  body force on this stage is hard to  validate. The results of the test on the formation of streaks are shown in figure \ref{Po-energy-Tf}($a$). When we target the first streak formation stage  $(\beta=0.35$ only when $ t\in(8,23))$, 
     laminarisation was found to fail. When we targeted the first two streak formation stages ($\beta=0.35$ only when $t\in(8,23) \& (38-51)$),
     the laminarisation was observed. 
     When targeting the first streak formation stage, although laminarisation failed, a clear drop of streamwise vorticity can be observed.  When targeting on first two streak formation stages, the streamwise vorticity can still rise again, but finally decay. The results (sky blue and red lines in figure \ref{Po-energy-Tf}(\textit{a}))  are quite consistent with the orange line in figure \ref{Po-energy-beta}$(b)$. In contrast, targeting the stage of the regeneration of streamwise vortices, as seen in figure \ref{Po-energy-Tf}$(b)$, seems to make less difference in the flow, although the flow is not periodic any more. Even when we impose the body force for three regeneration stages, no laminarisation happens. Therefore, we can now firmly conclude that it is the formation of streaks that is altered by the body force, leading to the more stable streaks and a collapse of turbulence. 
     
     Further, we think the key point for laminarisation is the flattening of the forced laminar profile, not the acceleration of flow near the wall, given that the nonlinear optimisations at $Re=2400$ and $Re=3000$ aimed to reduce the force near the wall rather than in the centre (as was shown in figure \ref{forceandUz}).
     The acceleration of flow near the wall is just a result of the fixed flux condition in our calculations. In the experiment of K\"uhnen $\textit{et al.,}$ \cite{kuhnen2018destabilizing}, the choice of accelerating the flow near the wall may be because it is more operable. 

  \subsection{Linear versus nonlinear optimisation}
     Through the construction of a suitable objective function, variational techniques can be used to find a theoretical optimal for laminarisation. In addition, the results of the nonlinear calculations can give insights into physics. 
     Through the optimisation, we find the minimal force to laminarise the turbulence, and confirm that the key to laminarisation is damping the flow in the centre, 
     which must be more efficient
     than accelerating the flow near the wall. 
     At higher Reynolds numbers,
     a more flattened forced laminar profile is required to laminarise turbulence. 
     
     After nonlinear optimisation,
     the net power saving  after laminarisation is $26.7\%$ at $Re=2400$ and $27.3\%$ at $Re=3000$ (the calculation for net power saving is discussed in Appendix A.1), while Song's unoptimised force \citep{song2014direct} already achieves
     $23.1\%$ at $Re=2400$ and $26.1\%$ at $Re=3000$.
    The reason 
    why the improvement is only small
    is that the extra pressure gradient term $ \frac{4}{Re}\beta$ (see equation \eqref{extrap}) for fixing mass flux increases after our optimisation, because the force aims to suppress the flow in the central region.
Although the amplitude of the final optimal force is reduced a lot compared to Song's input force \citep{song2014direct}, the total force including the extra pressure gradient $ \frac{4}{Re}\beta$ is not reduced significantly. 
  
     
In general, nonlinear optimisation for laminarisation is difficult, compared to the problem of triggering turbulence. 
Difficulties include the need to optimise
over several turbulent initial conditions,
and the sensitivity to initial conditions
of the turbulent state itself.
At high Reynolds numbers,  a large force might be necessary to pull the flow back to a laminar state,  but perhaps this could be offset to some extent by spatial localisation of the force. However, we have not observed a localised body force for laminarisation. The absence of azimuthal localised body force is quite reasonable, as a patch of turbulence can appear at any azimuthal position, thus the force needs to work on all angles.  
     For streamwise localisation, for the current length of the computational domain, we did not find localisation prior to imposing the filter in preliminary calculations.
    We expect that a localised optimal should exist for a sufficiently long pipe, but it is beyond the scope of this work.
     
     For the linear optimisation, it is computationally much quicker than the nonlinear case, and it is easy to constraint a zero mass flux upon the optimal forced laminar velocity,  then the corresponding body force does not cause an increase or decrease of mass flux, i.e. no change of extra pressure term $ \frac{4}{Re}\beta$.  In the current calculation, we used a longer target time $T=600$ to check whether the turbulence laminarise, thus a smaller force is allowed to laminarise turbulence. As a result, we can obtain a net power saving of $32\%$ at Re=2400 and $27\%$ at Re=3000. However, this method is also hard to extend to high Reynolds numbers.  For an increase of Reynolds number, the TG of more modes also needs to be reduced, but the number of modes is not known a priori, nor the reduction required for each.  
     
     While we have extended the optimisation
     for laminarisation to active forces (not purely damping), 
     the time-independence of the force in our calculations is different from that 
     employed in
     many classic flow control strategies, 
     which target directly or indirectly the suppression of streamwise vortices to reduce turbulence and drag. 
     Considering time-varying active forces 
     might achieve a more promising net power saving. 

   
\begin{acknowledgments}
 This work used the Cirrus UK National Tier-2 HPC Service at EPCC (http://www.cirrus.ac.uk) funded by the University of Edinburgh and EPSRC (EP/P020267/1).
\end{acknowledgments}

\appendix

\section{The calculation of net power saving}
   Suppose that the force rapidly induces laminarisation from 
   the turbulent state, while after laminarisation, it must be constantly maintained to suppress the onset
   of turbulence due to small perturbations.  This simplifies 
   calculation of the power saving by allowing us to calculate
   only the forces applied to the final laminar state.


   As the body force both accelerates and decelerates the flow,
   it could be decomposed into `active' and `passive' components
   \begin{equation}
  \vec{F}=\vec{F}_{active}+\vec{F}_{passive},
  \end{equation}
  where each is non-zero only where $\vec{F}\cdot\vec{u}$ is 
  positive or negative respectively.
  If we suppose that the passive component could be implemented by some means requiring no direct energy input, e.g.\ damping by a porous medium,
  then the energy cost is given by
  \begin{equation}
      W_{cost} = W_{driving} + W_{active},
  \end{equation}
  where `driving' refers to work done by the background pressure gradient.  In energy balance, we have that 
\begin{equation}
      W_{driving} + W_{active} = W_{passive} + W_{dissp.}\,
  \end{equation}
  where $W_{driving}=\left< \frac{4(1+\beta)}{Re}u_{tot,z}\right>$
  and $W_{dissip.}=\left< \frac{1}{Re}(\boldsymbol{\nabla}\times \vec{u}_{tot})^2\right>$.
  Given this balance, we may rewrite
  \begin{equation}
      W_{cost} = (W_{driving} + W_{active} + W_{passive} + W_{dissip.})/2
  \end{equation}
  where $W_{active} + W_{passive}=\langle |\vec{F}\cdot\vec{u}|\rangle$
  is computationally a little more straight forward to calculate than 
  $W_{active}$ and $W_{passive}$ individually.


   The net-power saving can be calculated as following,
   \begin{equation}
   	\mathcal{PR}_{lam/tur}=(W_{tur}-W_{lam})/W_{tur},
   \end{equation}
   where $W_{tur}$ represents the energy cost of flow without body force, $W_{lam}$ is the energy cost of forced laminar flow.

\bibliography{apssamp}

\end{document}